\documentclass[aps,showpacs,twocolumn,showkeys,superscriptaddress,amssymb,amsmath,reprint]{revtex4-1}
\usepackage{epsfig,amsopn}
\usepackage{graphicx}
\usepackage{wrapfig}
\usepackage{color}
\usepackage{amsmath,amssymb}
\usepackage{enumerate}
\usepackage{verbatim}
\usepackage{bm}
\newcommand\bea{\begin{eqnarray}}
\newcommand\eea{\end{eqnarray}}
\newcommand\beq{\begin{equation}}
\newcommand\eeq{\end{equation}}

\def\nn{\nonumber}
\def\f{\frac}

\def\ep{\epsilon}

\def\De{\Delta}

\DeclareMathOperator{\sgn}{sgn}

\begin{document}
\title{Transport in a long-range Kitaev ladder: role of Majorana and subgap Andreev states} 
 \author{Ritu Nehra}
 \email{ritunehra@iiserb.ac.in}
 \author{Auditya Sharma} 
 \email{auditya@iiserb.ac.in}
 \affiliation{ Department of Physics, Indian Institute of Science Education and Research, Bhopal 462066, India.}
 \author{Abhiram Soori}
\email{abhirams@uohyd.ac.in}
\affiliation{ School of Physics, University of Hyderabad, C.R. Rao Road, Gachibowli, Hyderabad 500046,  India.}

\begin{abstract}  
We study local and non-local transport across a two-leg long-range
Kitaev ladder connected to two normal metal leads. We
focus on the role of the constituent Majorana fermions and the subgap Andreev
states. The double degeneracy of Majorana fermions of the individual
legs of the ladder gets lifted by a coupling between the two leading
to the formation of Andreev bound states. The coupling can be induced
by a superconducting phase difference between the two legs of the
ladder accompanied by a finite inter-leg hopping. Andreev bound states
formed strongly enhance local Andreev reflection.  When the ladder and
normal metal are weakly coupled, the Andreev bound states, which are the controlling factor, result in weak
nonlocal scattering. In sharp contrast, when the ladder - normal metal 
interface is transparent to electron flow, we find that the subgap Andreev
states enhance nonlocal conductance strongly. The features in the local and nonlocal conductances resemble the spectrum of the isolated
ladder. Long-range pairing helps lift the degeneracy of the Majorana
modes, makes them less localized, and thus inhibits local transport,
while aiding non-local transport. In particular, long-range pairing
alone (without a superconducting phase difference) can enhance crossed Andreev reflection.
\end{abstract}
\maketitle

\section{Introduction} 
Majorana fermions~\cite{franz2010race,beenakker2013search,aguado17} in
condensed matter systems have received a great deal of attention in
the last couple of decades, with the Kitaev chain~\cite{kitaev2001unpaired}
proving to be the standard model. These exotic particles, which
may be conceptualized as special linear
combinations of electrons and holes, have been realized experimentally
in semiconductor quantum wires, exploiting the interplay of spin-orbit 
coupling, superconductivity and Zeeman field effects~\cite{lutchyn2010majorana,
oreg2010helical,mourik2012,Das2012,albrecht2016,Zhang2018}.
The promise of a \emph{robust} topological quantum computer~\cite{nayak08}
which has propelled a number of experiments on the one hand, and
the elegant theoretical ideas~\cite{beenakker2015random} involved on the
other, have been the strong motivations that have driven research on
Majorana fermions. 

The effect of Majorana fermions on transport is a topic of
considerable interest
~\cite{wakatsuki2014majorana,ioselevich2011anomalous,badiane2011nonequilibrium}.
Transport across a superconducting system connected to metallic leads
on either side, maybe classified into two types: local and
non-local. Nonlocal transport in such systems is mediated by two
processes: electron tunneling~(ET) and crossed Andreev
reflection~(CAR).  Electron tunneling (crossed Andreev reflection) is characterized by an electron incident from one normal metal resulting in an electron (a hole)
emitted in the other normal metal.  But, often the currents carried by
ET and CAR almost cancel out and a conductance measurement cannot
probe non-local transport~\cite{nilsson2008, soori19trans}. On the
other hand, local transport in any normal metal superconductor
interface is mediated by Andreev reflection~(AR) and electron
reflection~(ER). The incident electron from the normal metal reflects
back as a hole in the former, while it reflects back as an electron in the latter. Local
transport at normal metal superconductor interface has been studied
both theoretically and
experimentally~\cite{kseng01,mourik2012,Das2012,
  albrecht2016,Zhang2018,dassarma12,lin2012}.

In the topological phase of the Kitaev chain, each end carries a
Majorana fermion. When the two ends are connected to metallic leads,
it is conceivable that the edge Majorana fermions may jointly
contribute to the transport through the chain.  Recent
work~\cite{atala2014observation,hugel2014chiral,sun2016topological,
wang2016flux,nehra2018many,soori17,nehra19}
has shown that the ladder geometry encapsulates rich physics that
maybe exploited for a variety of purposes. In particular, when the
Kitaev chain is replaced by a Kitaev ladder, while still retaining the
unique topological properties, an enhancement of crossed Andreev
reflection under suitable conditions maybe
engineered~\cite{soori17,nehra19}. The Kitaev ladder system is
characterized by two Majorana fermions at each end, as opposed to just
one in the Kitaev chain. As expected, the superconducting term ensures
that the spectrum is gapped. A finite coupling strength along with
superconducting phase difference between the legs of the ladder
results in two effects.  On the one hand it induces plane wave states
within the superconducting gap known as subgap Andreev states, and by
suitably controlling them crossed Andreev reflection may be
enhanced~\cite{soori17,nehra19}. On the other hand, when the ladder is
tuned within the topological phase, the Majorana fermions at each end
become bound.  As the inter-leg coupling strength is increased maintaining 
a finite superconducting phase difference between the two legs, a
splitting of energy levels (of the order of inter-leg coupling
strength) between the two Majorana bound pairs is observed. Thus,
while the subgap Andreev states enhance non-local transport, the
presence of Majorana fermions enhances local AR. However, the
underlying controlling factors for these contrasting features are the
same: the inter-leg coupling and superconducting phase difference. The
current work is dedicated to a study of the effect of these competing
processes in a junction made up of 
\begin{figure*}
\centering
\includegraphics[scale=1.0]{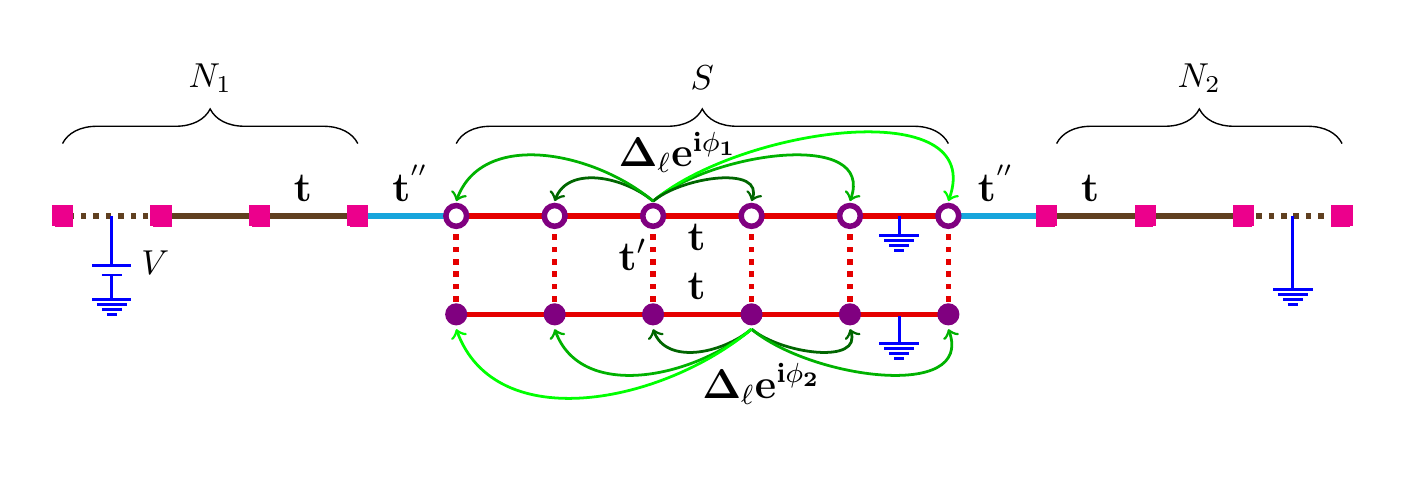}
\caption{ The long range superconducting Kitaev ladder~($S$) is
  connected to two normal metal leads $N_1$ and $N_2$.  Each of $N_1$
  and $N_2$ is modeled by a one-dimensional tight binding model with a
  hopping amplitude $t$ and a chemical potential $\mu$. $S$ corresponds to
  a lattice model as shown and has two coupled parallel long range
  Kitaev chains. The intra-chain hopping is set to be $t$ and
  inter-chain hopping is given by $t'$ along with on-site chemical
  potential $\mu$. Each site of each Kitaev chain is connected to
  every other site in the same Kitaev chain by long range superconducting pairing amplitude 
  $\Delta_l  e^{i\phi_{\sigma}}=\Delta l^{-\alpha} e^{i\phi_{\sigma}}$ (with long
  range parameter $\alpha$ where $l$ is the distance between the connecting sites)
  along with superconducting phase factor $\phi_{\sigma}$ in
  $\sigma^{th}$ leg of the ladder. The normal metal leads are attached
  to the upper leg($\sigma=1$) of the ladder with a hopping strength
  $t^{\prime\prime}$. A voltage bias $V$ is applied from N$_1$,
  keeping S and N$_2$ grounded.}
\label{fig:01}
\end{figure*}

 metal, Kitaev ladder and metal.

The rest of the paper is organized as follows. In
the next section we introduce the model through a Hamiltonian and
discuss the dispersion and the appearance of Majorana modes. The following section describes the transport properties of the long-range Kitaev ladder 
when it is connected to metallic 
leads, in relation to the properties of the long-range Kitaev ladder. The concluding section summarizes
the central findings.
\section{Long Range Kitaev ladder}

\subsection{Hamiltonian}
The long-range Kitaev ladder~\citep{bhattacharya2019critical,ares2018entanglement,PhysRevB.97.155113}
is a system of two parallel spinless
p-wave superconductor wires coupled to each other. To study transport
across this system, it is useful to consider a three-terminal setup of
normal metal-superconductor-normal metal~($N_1SN_2$) such that the left
metal~($N_1$) is maintained at a voltage $V$ while the
superconductor~($S$), and the right metal~($N_2$) are grounded. The
superconductor of the setup is hence connected to two normal
metals~($N_1,N_2$) with coupling strength $t''$ as shown in
Fig.~\ref{fig:01}.

The Hamiltonian for the setup is given by
\begin{equation}
\label{eq:fullham}
H=H_{N_1} + H_{N_1 S} + H_S + H_{N_2 S} + H_{N_2},
\end{equation}
where
\begin{align}
\label{eq:1}H_{N_1}=&-t\displaystyle\sum_{j\le 0}(c_{j+1}^{\dagger}c_{j}+c_{j}^{\dagger}c_{j+1})
-\mu\displaystyle\sum_{j\le 0}c_{j}^{\dagger}c_{j}\\
\label{eq:2} H_{N_2}=&-t\displaystyle\sum_{j\ge L+1}(c_{j+1}^{\dagger}c_{j}+c_{j}^{\dagger}c_{j+1})
-\mu\displaystyle\sum_{j\ge L+1}c_{j}^{\dagger}c_{j},
\end{align}
are the Hamiltonians for the normal metal leads, 
\begin{align}
H_{N_1 S}=& -t'' [c_{0}^{\dagger}c_{1,1}+h.c.],\\
H_{N_2 S}=& -t'' [c_{L+1}^{\dagger}c_{L,1}+h.c.] 
\end{align}
are the Hamiltonians that connect the leads to the superconductor, and
\begin{widetext}
\begin{align}
\label{eq:hs}
H_{S}=-&\displaystyle\sum_{j=1\atop \sigma=1,2}^{L-1}(tc_{j+1,\sigma}^{\dagger}c_{j,\sigma}
+t^{\prime}c_{j,\sigma}^{\dagger}c_{j,\bar\sigma}+h.c.)
-\displaystyle\sum_{j=1\atop \sigma=1,2}^{L} \mu\Big(c_{j,\sigma}^{\dagger}c_{j,\sigma}-\frac{1}{2}\Big)
-\displaystyle\sum_{j=1\atop \sigma=1,2}^{L-1}\Bigg[\displaystyle\sum_{j'=j+1}^{L}\Bigg(\frac{\Delta}{(j'-j)^{\alpha}}
e^{i\phi_{\sigma}}c_{j,\sigma}^{\dagger}c_{j',\sigma}^{\dagger}+h.c.\Bigg)\Bigg]
\end{align}
\end{widetext}
corresponds to the long-range Kitaev ladder. Here,
$c^{\dagger}_{j,\sigma}(c_{j,\sigma})$ are fermion creation
(annihilation) operators on $j^{th}$ site ($j=1,2,...,L$) of the
$\sigma^{th}$ leg ($\sigma,\bar\sigma=1,2$) of the ladder with $a=1$
as unit cell length. The fermion creation (annihilation) operators
for normal metal leads are given by $c^{\dagger}_j~(c_j)$ where $j$~is
a site on the left metal~(N$_1$) for $j\le0$ and on the right
metal~($N_2$) for $j\ge L+1$. The hopping along the legs of the Kitaev
ladder as well as in the normal metal leads are maintained at $t$ with
onsite chemical potential $\mu$. The inter-leg hopping $t'$ couples
the two legs of the Kitaev ladder. $\Delta_l e^{i\phi_{\sigma}}$ is
the 
\begin{figure*}
 \includegraphics[scale=0.33]{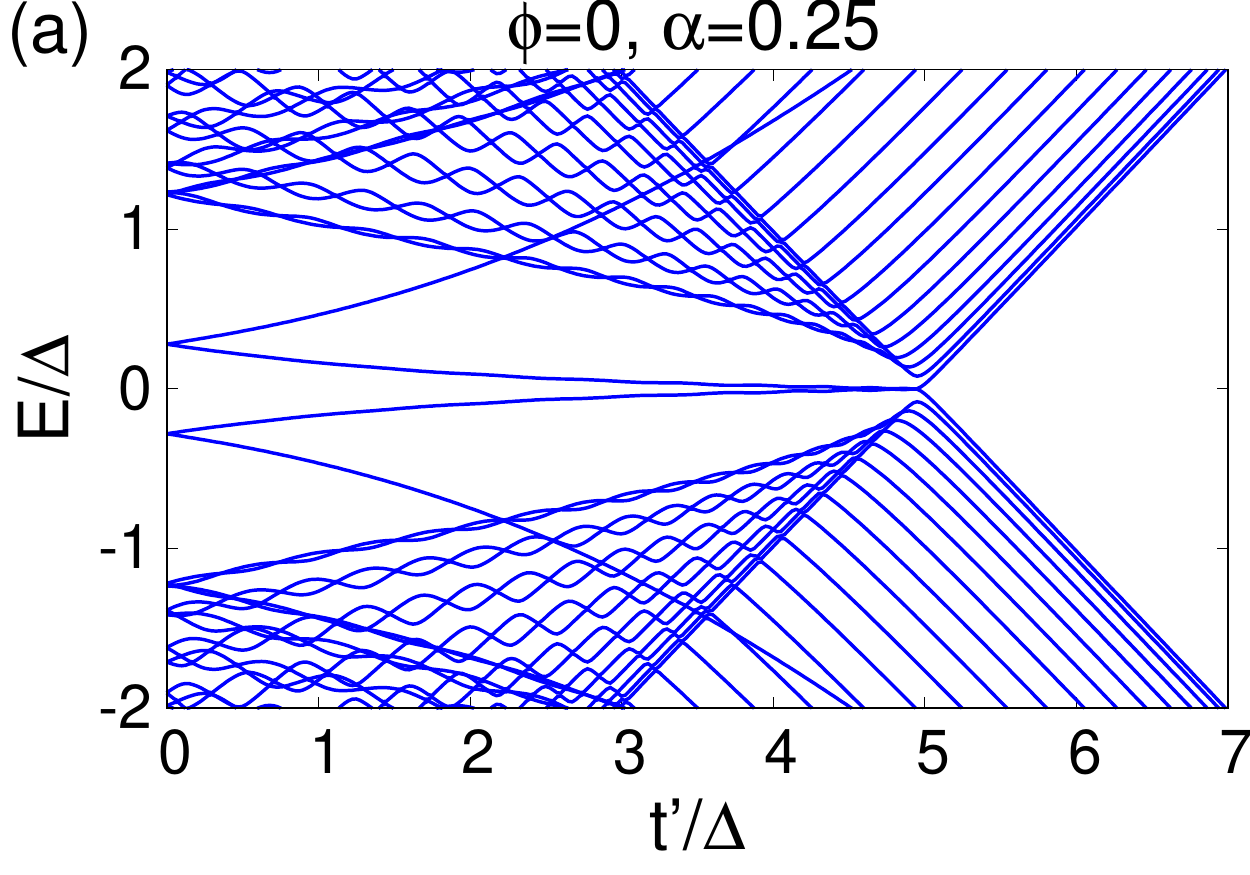}
 \includegraphics[scale=0.33]{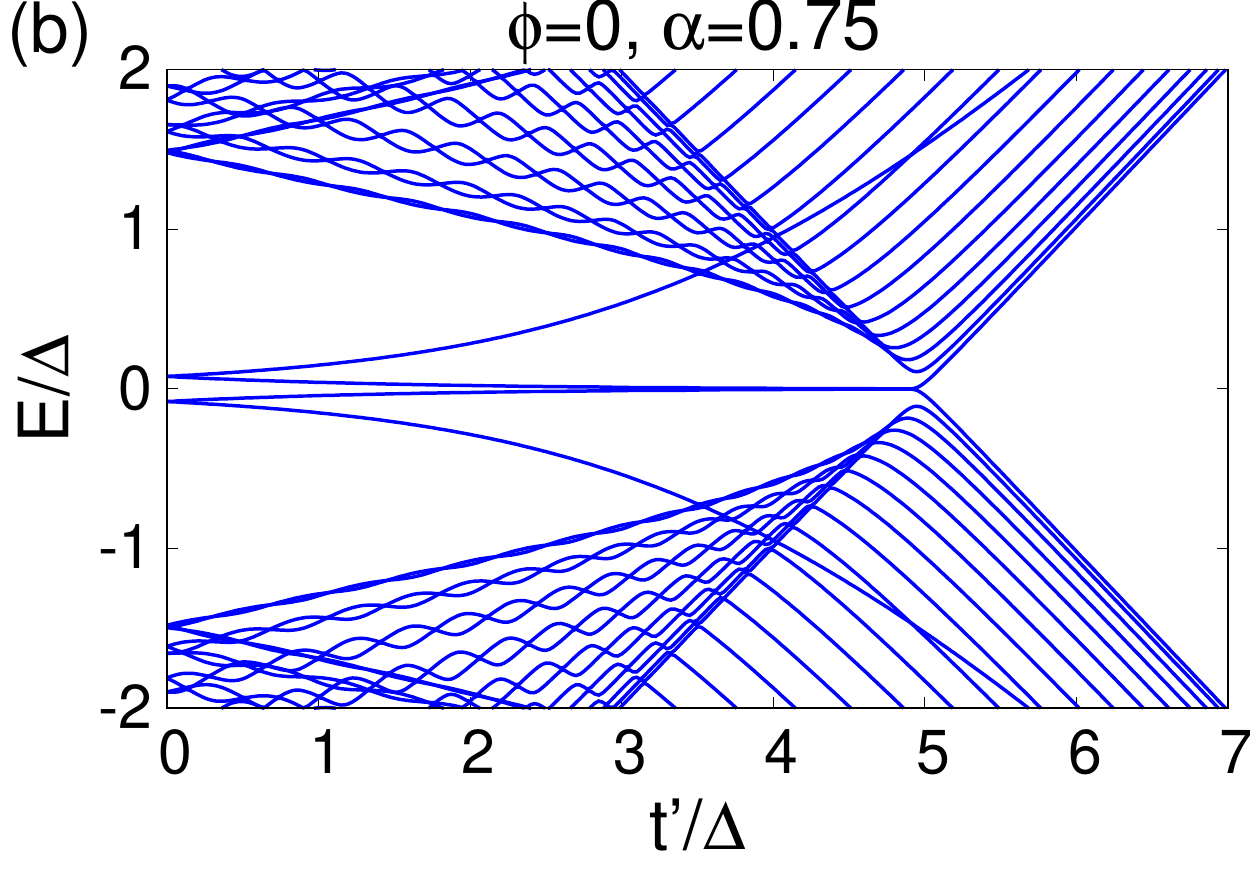}
 \includegraphics[scale=0.33]{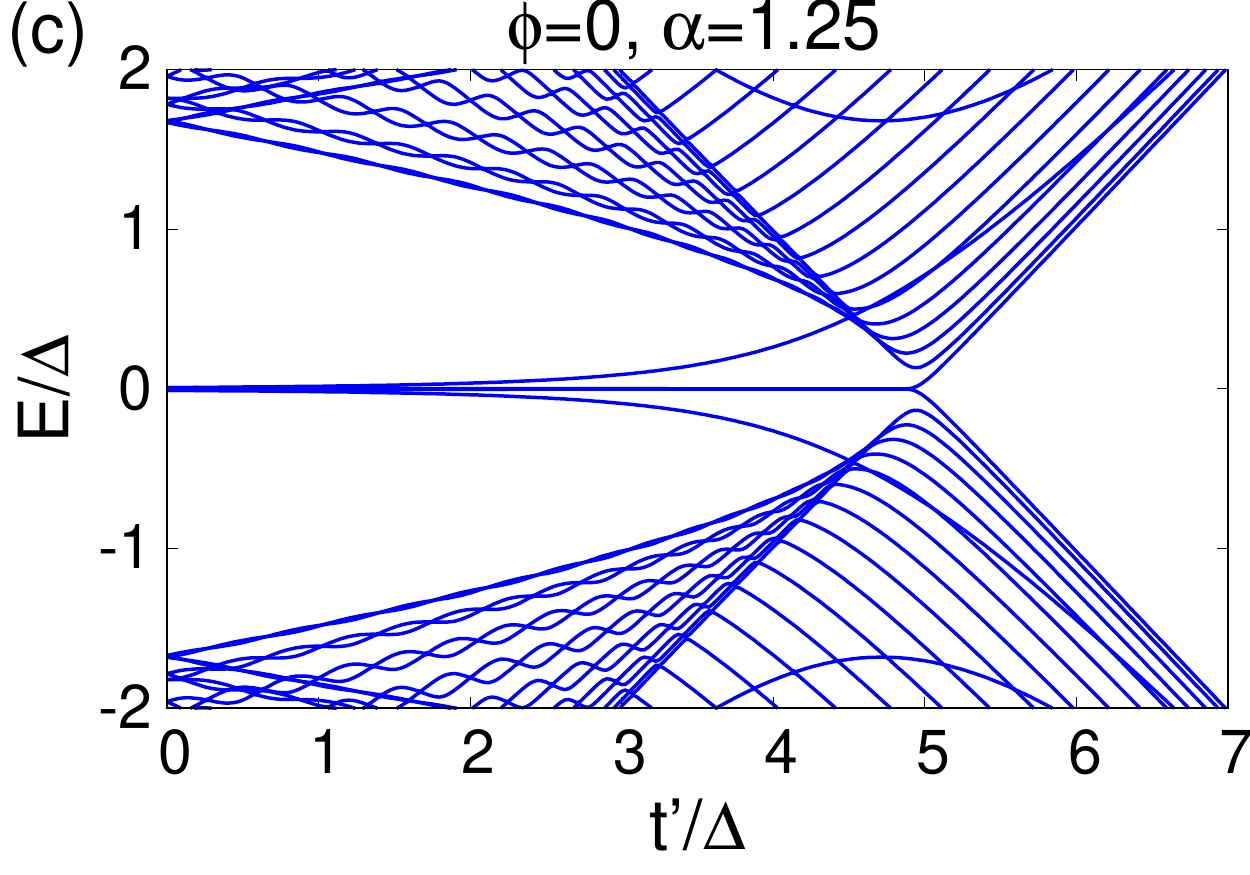}
 \includegraphics[scale=0.33]{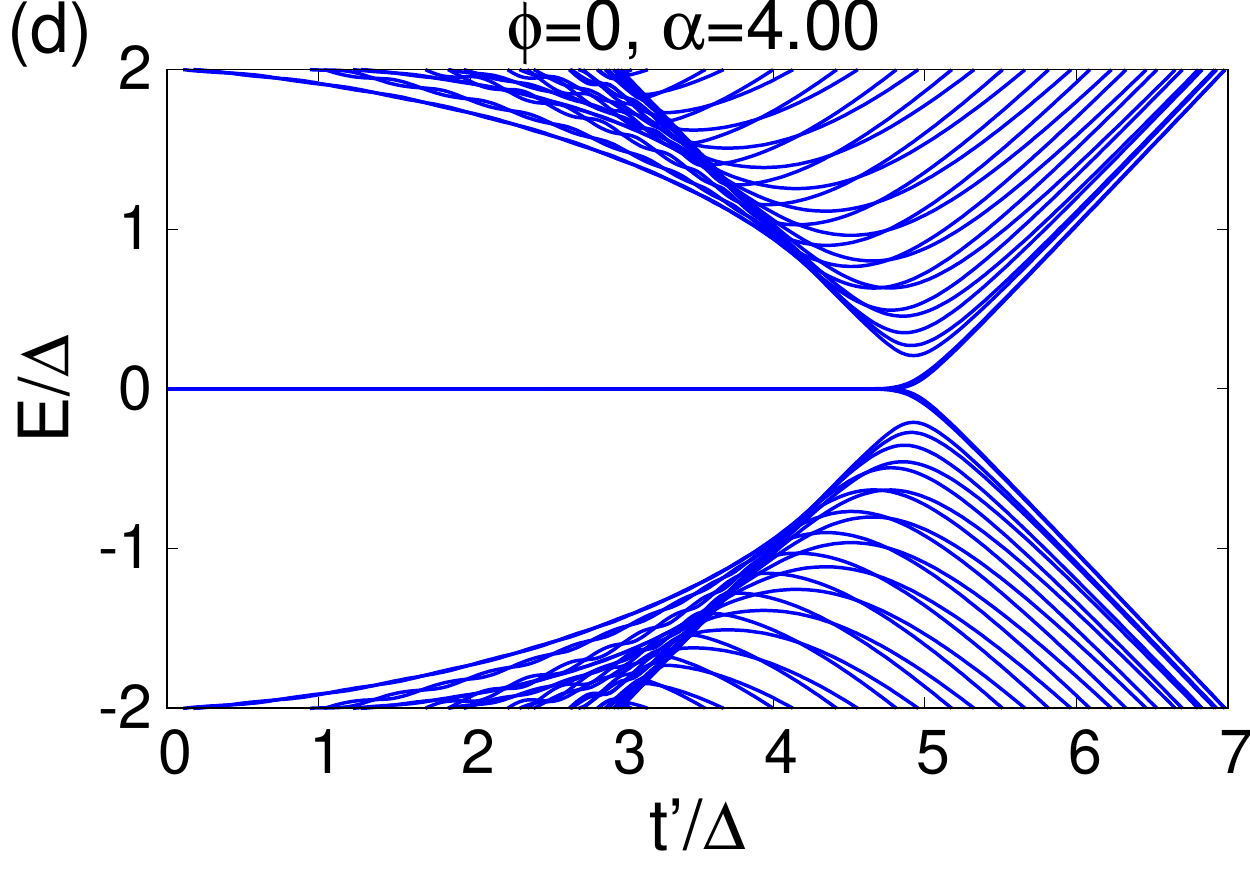}\\
 \includegraphics[scale=0.33]{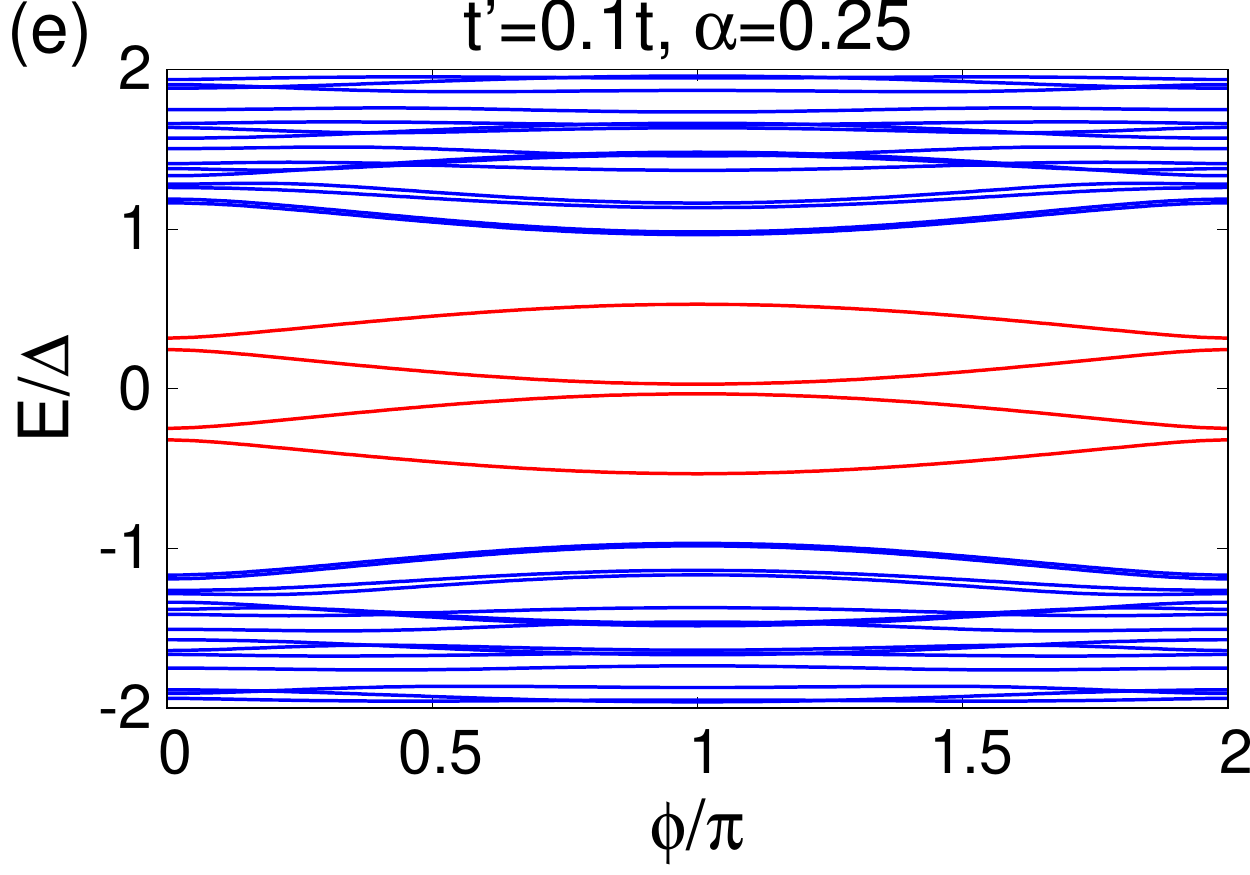}
 \includegraphics[scale=0.33]{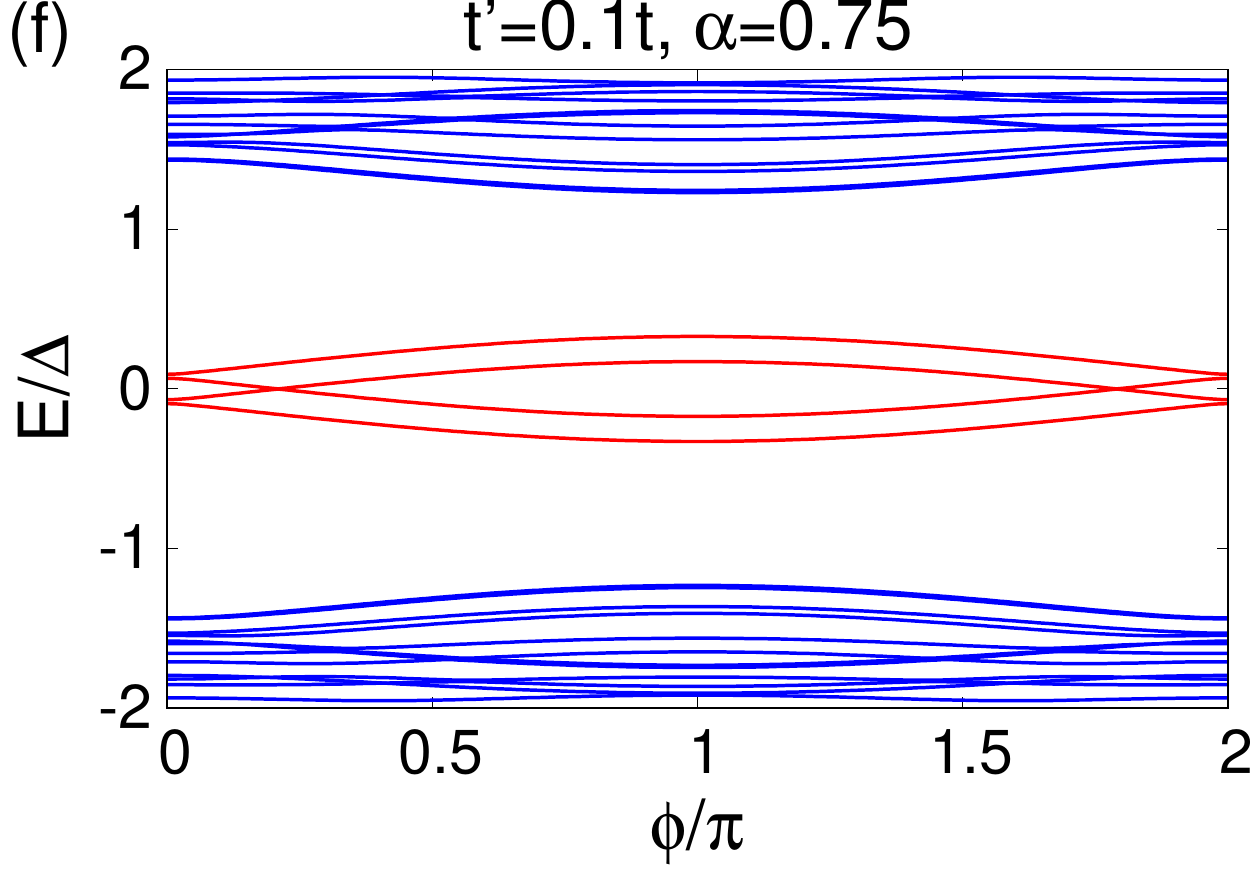}
 \includegraphics[scale=0.33]{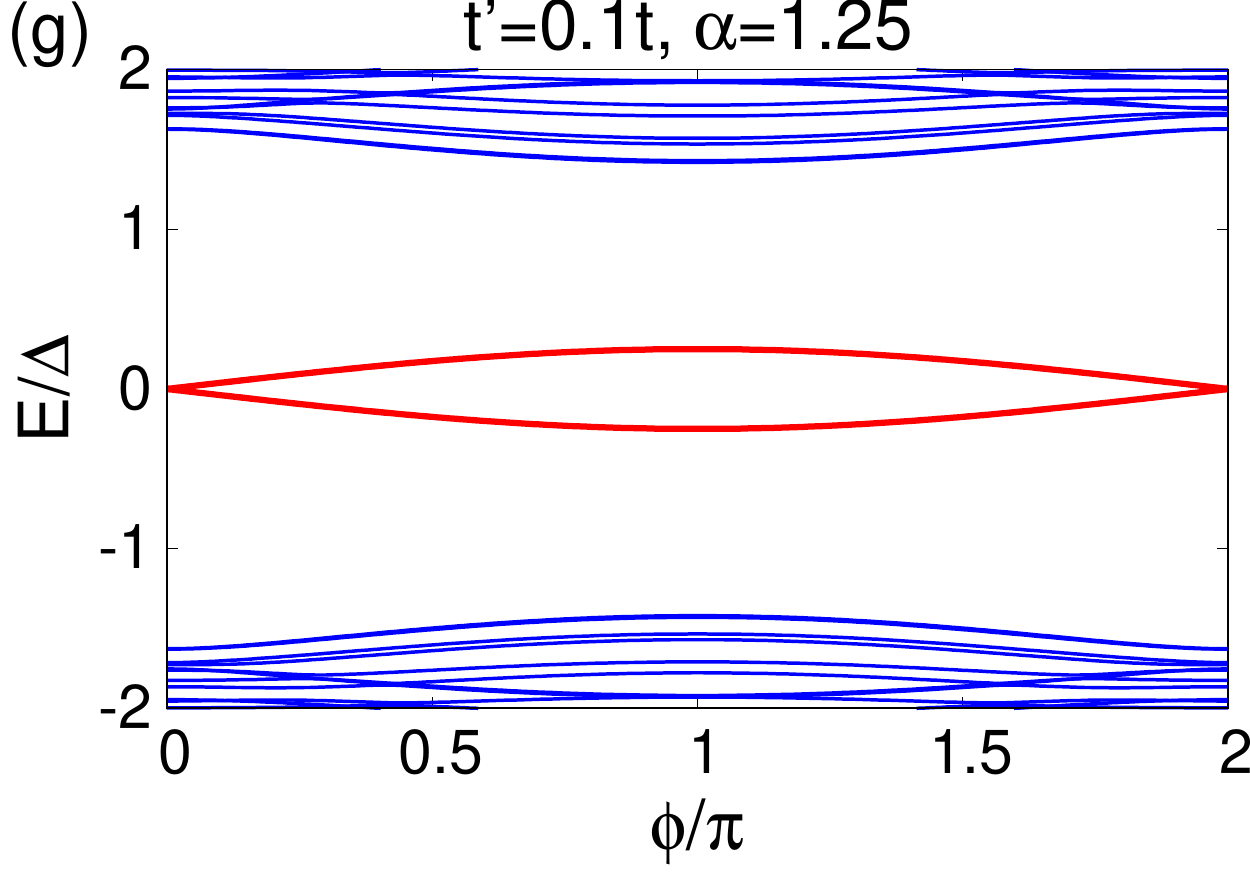}
 \includegraphics[scale=0.33]{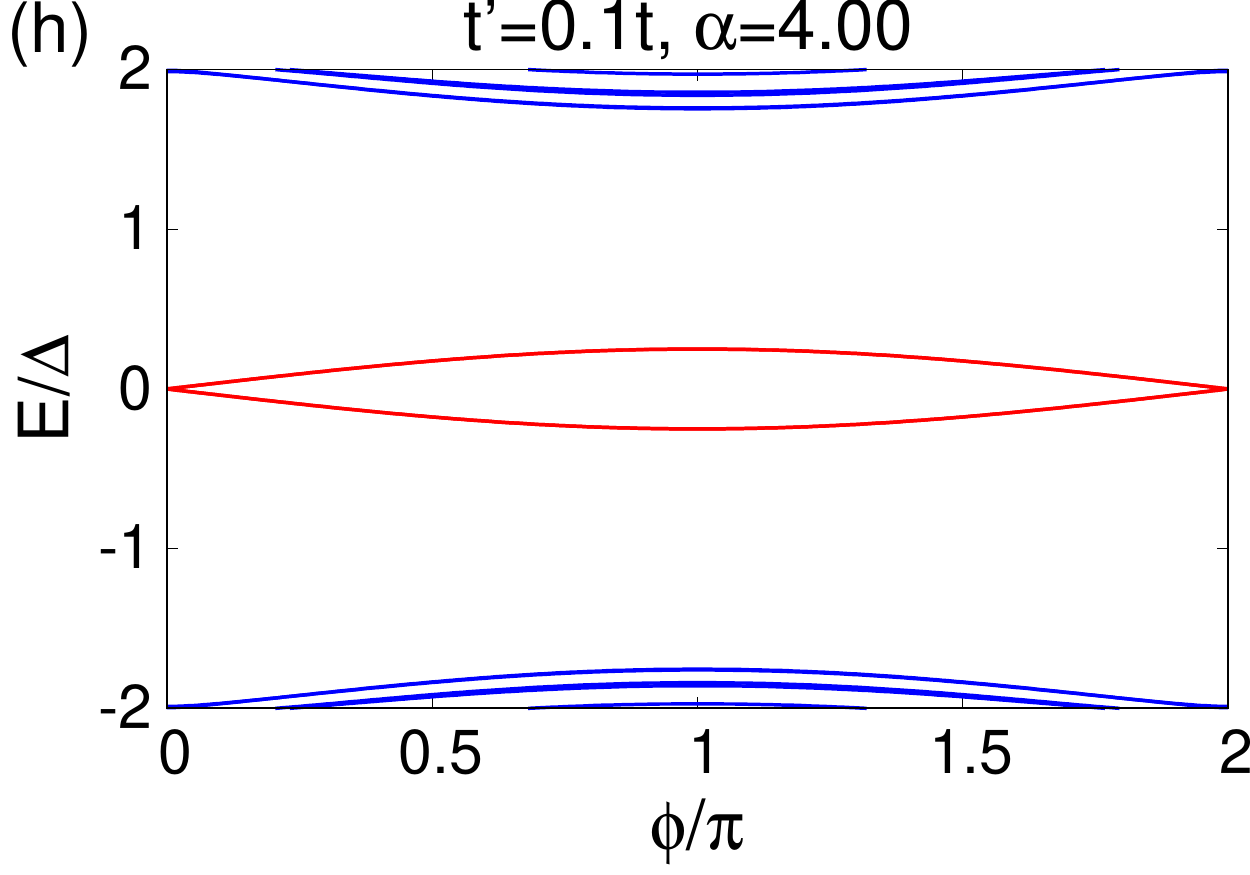}
 \caption{The subgap energy spectrum of the long-range Kitaev ladder
   with open boundary conditions as a function of (a-d) the inter-leg
   hopping $t'$ with $\phi=0$, and (e-h) the superconducting phase
   difference ($\phi$) with $t'=0.1t$.  The other set of parameters
   are taken to be $\De=0.4t$, $\mu=0$ and $L=40$ with long range
   parameter (a,e) $\alpha=0.25$, (b,f) $\alpha=0.75$,
   (c,g) $\alpha=1.25$, (d,h) $\alpha=4.00$. The long-range parameter
   $\alpha\rightarrow\infty$ i.e. large $\alpha$ recovers the short
   range Kitaev ladder as shown in (d,h). For the lower values of
   $\alpha$ the Majorana modes split apart as shown in (a-c,e-f). The
   degeneracy of Majorana zero modes shown by red lines is lifted by
   breaking of symmetries due to the presence of superconducting phase
   difference ($\phi$) as well as long-range parameter ($\alpha$). }
 \label{fig:02}
\end{figure*}
superconducting pair potential with a constant superconducting phase
$\phi_{\sigma}$ in $\sigma^{th}$ leg of the ladder. In this work, we
consider a long-range distance-dependent superconducting pair
potential~\citep{bhattacharya2019critical} $\Delta_l=\Delta
l^{-\alpha}$ that connects each site of the Kitaev ladder to other
sites (in the same leg) as shown in Fig.~\ref{fig:01} by green
curved lines and decays along the legs of the ladder with distance
between two connecting sites ($l=|j-j'|$) as a power-law
($\frac{1}{l^\alpha}$) with exponent $\alpha>0$. Open boundary
conditions, which are natural for the current setup that involves
transport from and into metallic leads, are assumed. The contact
between metal leads and Kitaev ladder is taken to have a tunable
coupling strength $t''$.

\subsection{Dispersion}
In order to study the dispersion and topological properties of the
long-range Kitaev ladder, it is convenient to use the Majorana
basis~\citep{degottardi2011topological}. The
fermion operator splits into two Majorana fermion operators:
\begin{equation}
\label{eq:ut}c_{j,\sigma}=\frac{1}{2}(\gamma^A_{j,\sigma}+i\gamma^B_{j,\sigma}),\;
c^{\dagger}_{j,\sigma}=\frac{1}{2}(\gamma^A_{j,\sigma}-i\gamma^B_{j,\sigma})
\end{equation}
where $j$ is the site index of the ladder, $\sigma$ is the leg index
of the ladder and $A,B$ are two types of Majorana fermions. In order to
maintain the anti-commutator relation of fermions, these Majorana
operators should obey the following relations:
\begin{equation}
\gamma^{\eta}_{j,\sigma}=(\gamma^{\eta}_{j,\sigma})^{\dagger},
\;\lbrace\gamma^{\eta}_{i,\sigma},\gamma^{\eta'}_{j,\sigma'}\rbrace
=2\delta_{i,j}\delta_{\eta,\eta'}\delta_{\sigma,\sigma'}
\end{equation}
where $\eta, \eta'=A,B$, and $\sigma, \sigma'=1,2$.
The invocation of this unitary transformation in Eq.(\ref{eq:hs}) leads to
\begin{widetext}
\begin{align}
\label{eq:maj}H_{S}=-\frac{i}{2}\displaystyle\sum_{j=1\atop\sigma=1,2}^{L-1}
\Big[t(\gamma^{A}_{j+1,\sigma}\gamma^{B}_{j,\sigma}-\gamma^{B}_{j+1,
\sigma}\gamma^{A}_{j,\sigma})\Big]&-\frac{it'}{2}\displaystyle\sum_{j=1}^{L}
(\gamma^{A}_{j,1}\gamma^{B}_{j,2}+\gamma^{B}_{j,2}\gamma^{A}_{j,1})
-\frac{i\mu}{2}\displaystyle\sum_{j=1\atop\sigma=1,2}^{L}(\gamma^{A}_{j,
\sigma}\gamma^{B}_{j,\sigma})\\\nonumber-\frac{i\Delta}{2}\displaystyle\sum_{j
=1\atop\sigma=1,2}^{L-1}\displaystyle\sum_{j'=j+1}^{L}\Bigg[\frac{\sin\phi_{\sigma}}
{(j'-j)^{\alpha}}&(\gamma^{A}_{j',\sigma}\gamma^{A}_{j,\sigma}+\gamma^{B}_{j,\sigma}\gamma^{B}_{j',\sigma})+\frac{\cos\phi_{\sigma}}{(j'-j)^{\alpha}}(\gamma^{A}_{j,\sigma}\gamma^{B}_{j',\sigma}+\gamma^{B}_{j,\sigma}\gamma^{A}_{j',\sigma})\Bigg].
\end{align}
\end{widetext}
Further, we take the limit of $L\to\infty$ for the dispersion relation
to make sense.  A Fourier transformation of the Hamiltonian (with the basis spinor chosen to be
$[\gamma^A_{k,1},\gamma^A_{k,2},\gamma^B_{k,1},\gamma^B_{k,2}]$) in Eq.(\ref{eq:maj})
results in
\begin{align}
\label{eq:hamil}
H_s(k)=
\begin{bmatrix}
Q_k &A_k\\
A^{\dagger}_k &-Q_k
\end{bmatrix}
\end{align}
where
\begin{align}
A_k=-\frac{i}{2}\begin{bmatrix}
(\epsilon_k-\Delta_k\cos\phi_1) &t'\\
t' &(\epsilon_k-\Delta_k\cos\phi_2)
\end{bmatrix},
\end{align}
\begin{align}
Q_k=-\frac{i}{2}\begin{bmatrix}
\Delta_k\sin\phi_1 &0\\
0 &\Delta_k\sin\phi_2
\end{bmatrix},
\end{align}
$\epsilon_k=(2t\cos ka+\mu)$, $\Delta_k=2i\Delta f_{\alpha}(k)$,
$f_{\alpha}(k)=\displaystyle\sum_{l=1}^{\infty}\frac{\sin
  (kal)}{l^{\alpha}}$, and $a$ is the lattice spacing. The dispersion relation of the isolated long range
Kitaev ladder Hamiltonian is given by
\begin{equation}
\label{eq:disp-ladder} 
E=\pm\sqrt{\epsilon_k^2+{t^{\prime}}^2+|\Delta_k|^2\pm 2
t^{\prime}\sqrt{\epsilon_k^2+|\Delta_k|^2~\sin^2 \frac{\phi}{2}}},
\end{equation}
where $\phi=\phi_1-\phi_2$ is the superconducting phase difference
between the two legs of the ladder. These phases can be managed by
superconducting quantum interference devices (SQUIDS)~\citep{PhysRevB.89.174514}.

The  spectrum of the long-range Kitaev ladder is depicted in
Fig.~\ref{fig:02} when open boundary conditions are imposed for
different $\alpha$. In the large $\alpha$ limit
i.e. $\alpha\rightarrow\infty$ the system behaves as a short-range
Kitaev ladder as shown in Fig.~\ref{fig:02}(d,h). The gap region of
the short-range Kitaev ladder in the absence of superconducting phase
difference ($\phi=0$) features four degenerate zero energy Majorana
modes in Fig.~\ref{fig:02}(d). These zero energy Majorana modes are
highly localized on the edges of the ladder as compared to any other
state of the system which shows dispersive behavior. Each leg of the
ladder hosts one Majorana fermion at each end. A finite length of the
ladder $L$ couples the Majorana fermions (MFs) at the two ends of the legs of the ladder
and they split in energy with a splitting proportional to
$e^{-L/l_0}$~\citep{kitaev2001unpaired}, where $l_0$ is the decay
length in the ladder. However, the presence of the superconducting phase
difference enhances the hybridization of Majorana
modes of the different legs of the ladder;
as a result, the degeneracy of these Majorana fermions is lifted by an amount
$E=\pm \tilde{t}\sin{\frac{\phi}{2}}$ as shown by red lines in Fig.~\ref{fig:02}(h), 
where $\tilde{t}\sim t'$ is the effective coupling between the  Majorana fermions 
of the individual legs of the ladder. These states formed by the hybridization of 
Majorana fermions are called Andreev bound states~(ABSs). The other 
eigenstates are bulk states of the system and are called subgap 
Andreev states, which have been shown to be important for
non-local transport~\cite{soori17,nehra19}. These subgap
Andreev states~(SASs) formed by the hybridization of the bulk
bands of the individual legs of the ladder are shown by blue lines in
Fig.~\ref{fig:02}(h).

The long-range nature of the pairings controlled by $\alpha$ has
important effects on both Andreev bound states as well as on subgap Andreev states. As $\alpha$ is 
decreased the system shifts into the long range
limit where every site of the ladder is connected to the all other
sites by the superconducting term. This connection enhances the decay
length of localized Majorana modes inside the bulk of the ladder. This
results in splitting of Majorana states due to the coupling of the 
Majorana states (Andreev bound states) at the two ends of the ladder as shown in Fig.~\ref{fig:02}(a)
(Fig.~\ref{fig:02}(e)). When $\alpha$ is very small and the system
is truly long-ranged, the degeneracy of the MFs (ABSs) is
completely lifted even for very tiny $t'$. Moreover, as the
pairings become more and more long-ranged, the subgap Andreev states have a tendency to close in into the gapped region as
illustrated in Fig.~\ref{fig:02}(a-c,e-g).  In between the extreme
short-range limit and the extreme long-range limit the
Majorana modes remain close to degenerate for small values of $t'$ as
shown in Fig.~\ref{fig:02}(d,c) and very strong inter-leg
coupling is needed to lift this degeneracy.
\subsection{Topological Properties} 
\begin{figure}[h!]
 \includegraphics[scale=0.55]{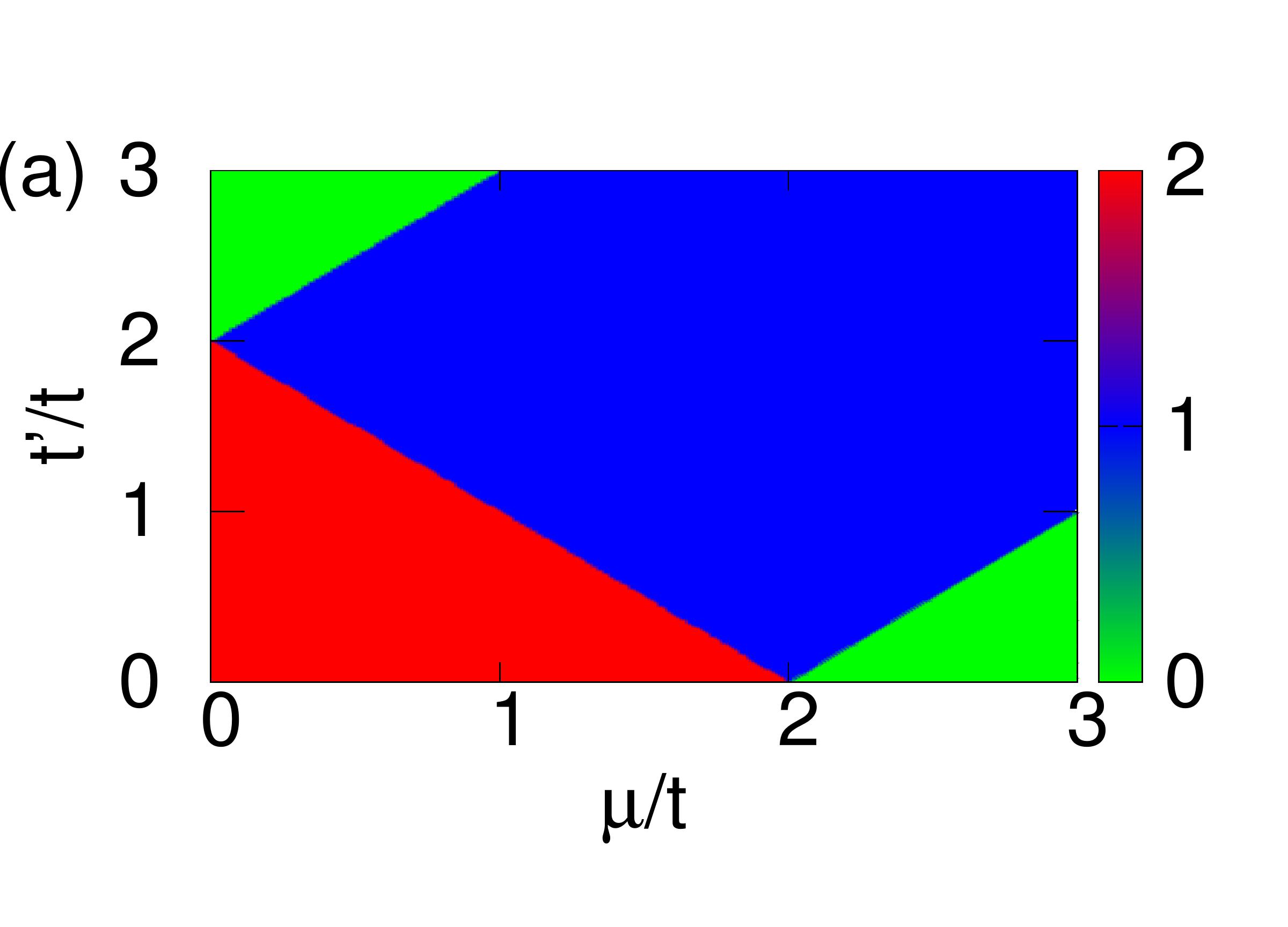}
 \includegraphics[scale=0.55]{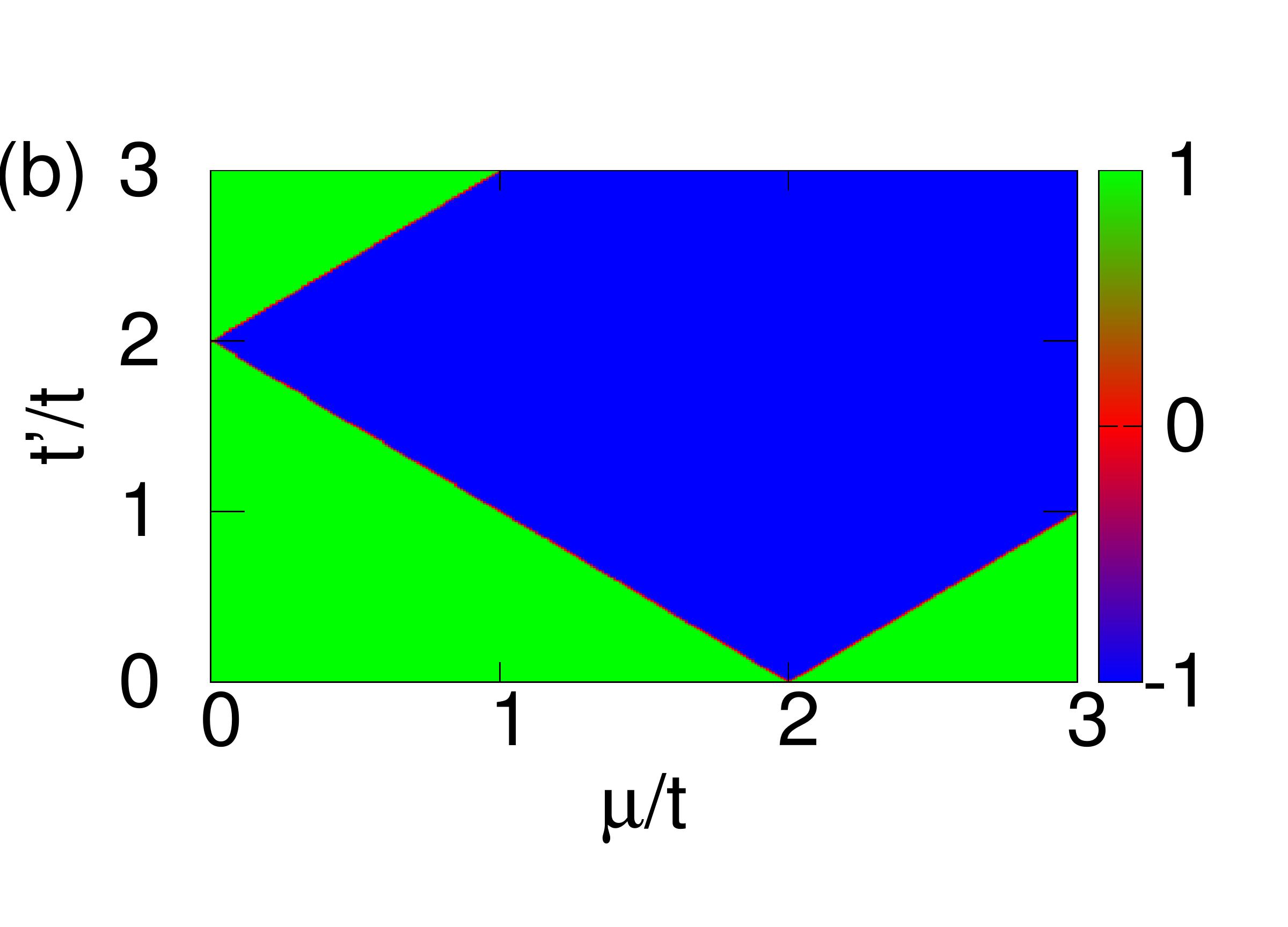}
 \caption{(a) The winding number ($w$ in Eq.~\ref{eq:winding}) for the Kitaev ladder in the
   limit $\phi=0$ with varying inter-leg hopping and on-site chemical
   potential $\mu$ with $t=\Delta$. (b) The Majorana number ($M$ in Eq.~\ref{eq:mnumber}) for
   the Kitaev ladder for $\phi\neq0$ with varying inter-leg hopping and
   on-site chemical potential $\mu$. The green region in both cases
   shows the non-topological phase and the blue region signifies the
   presence of at most one Majorana mode at each end of the ladder. The red
   region corresponds to the presence of at most two Majorana modes per
   edge.}~\label{fig:03}
\end{figure}

One of the distinguishing features of the Kitaev system is its
topological properties, and can help improve our understanding of the
localization domains of the Majorana modes. The Kitaev ladder in the
absence of superconducting phase has all three symmetries
i.e. particle-hole symmetry, chiral symmetry, and time-reversal
symmetry~\citep{kitaev2001unpaired}. A system following these symmetries belongs to the BDI class for which the number of zero energy modes is calculated by a $\mathbb{Z}$ value topological invariant winding number $w$.
 To calculate the winding
 number of the Kitaev ladder~\citep{Maiellaro2018,PhysRevB.95.195160} with $\phi = 0$,
 the transformed Hamiltonian in Eq.(~\ref{eq:hamil}) is
 converted into off-diagonal form. The winding number ($w$)
 for such a system is given by
\begin{equation}
\label{eq:winding}
w=\frac{1}{2\pi i}\int_{-\pi}^{\pi}dk\Bigg(\frac{\partial\;\ln\;\det\;A_k}{\partial k}\Bigg).
\end{equation}
The winding number ($w$) for the case $\phi=0$ is represented in
Fig.~\ref{fig:03}(a). For $\phi=0$ the winding number of the system
takes the values $2,1,0$ depending on the choice of the system parameters. The maximum winding number
of $2$ corresponds to the case when two Majorana modes are present at each end of the
ladder. When a non-zero superconducting phase difference ($\phi\neq0$) is present, the Hamiltonian
breaks the time-reversal symmetry, and therefore the chiral symmetry as well. Therefore in
the presence of superconducting phase difference $\phi$ the system only possesses the
particle hole symmetry and thus falls in class D of the classification~\citep{PhysRevB.95.195160}. 
For class D, the $\mathbb{Z}_2$ value Majorana number($M$) is a 
good topological invariant which involves the calculation of the Hamiltonian's (Eq.~\ref{eq:hamil})
Pfaffian~\citep{kitaev2001unpaired,Maiellaro2018,WU20123530}: 
\begin{equation}
\label{eq:mnumber}
M = \sgn(\mathrm{Pf}[H_s(0)])\sgn(\mathrm{Pf}[H_s(\pi)]).
\end{equation}
The Majorana number takes the values $M=1$, and $M=-1$ for
non-topological and topological phase, respectively. The Majorana
number for the system with $\phi\neq 0$ is depicted in
Fig.~\ref{fig:03}(b) which shows a clear transition from
non-topological to topological phase as the different parameters of
the system are tuned. We see therefore that in this case, there will
be at most one Majorana fermion at the edge of the ladder. We emphasize
that the topological features of the long-range Kitaev ladder are
completely independent of the parameter $\alpha$.

\section{Transport}
We now study transport in this system with a focus on the role played
by the Majorana modes and subgap Andreev states, which have
contrasting tendencies. An incident electron must undergo one of four
different scattering processes: reflection as an electron, reflection
as a hole, transmission as an electron or transmission through the
system as a hole. These processes are called Electron reflection
(ER), Andreev reflection (AR), Electron tunneling (ET) and crossed
Andreev reflection (CAR), respectively.  Therefore, considering these
different processes, the wave function (on site indexed $j$) in $N_1$ and $N_2$ can be
written as $[\psi^e_j,~\psi^h_j]^T$ where 
\bea
\psi^e_{j}&=&e^{ik_eaj}+r_e e^{-ik_eaj} {~~\rm for~~} j \le
0\nn\\ \psi^h_{j}&=&r_he^{ik_haj} {~~\rm for~~} j \le 0\nn
\\ \psi^e_{j}&=&t_e e^{ik_eaj}{~~\rm for~~} j \ge L+1\nn\\
\label{eq:psi}
\psi^h_{j}&=&t_h e^{-ik_haj}{~~\rm for~~} j \ge L+1.
\eea
In the ladder region, the wave function has the form 
$[\psi^e_{j,\sigma},\psi^h_{j,\sigma}]^T$
with $1\leq j\leq L$ as the site index and $\sigma=1,2$ as the ladder leg index. 
At a given energy $E$,  $k_{e/h}a=\cos^{-1}\big[{(-\mu\mp E)}/{2t}\big]$.
The scattering amplitudes $r_e$, $r_h$, $t_e$, $t_h$ for ER, AR, ET and CAR respectively can be 
calculated from the equation of motion combined with the form
of wavefunction in Eq.~\eqref{eq:psi} as discussed in the appendix.
  
 The local conductance $G_{11}$ is the ratio of the differential change in current in N$_1$ to the differential change in applied voltage in N$_1$,  i.e., $G_{11}=\frac{dI_1}{dV_1}$. The nonlocal conductance (or transconductance) is the ratio of the differential change in
 the current in N$_2$ to the differential change in the applied voltage in N$_1$,  i.e., $G_{21}=\frac{dI_2}{dV_1}$.
 These conductances can be expressed in terms of various scattering
 probabilities using Landauer-B\"uttiker
formalism~\cite{landauer1957r,landauer1970r,buttiker1985m,buttiker1986m,datta1995} as
\begin{align}
G_{11}&=\f{e^2}{h}\Big[1-|r_e|^2+|r_h|^2\f{\sin{k_ha}}{\sin{k_ea}}\Big]\nn, \\
G_{21}&=\f{e^2}{h}\Big[|t_e|^2-|t_h|^2\f{\sin{k_ha}}{\sin{k_ea}}\Big] \label{eq:conductance}.
\end{align} 
As expected, the local processes ER and AR contribute towards local conductance whereas the 
nonlocal processes ET and CAR contribute towards nonlocal conductance. Further, dominance of 
AR over ER is signaled by $G_{11}>e^2/h$ while dominance of CAR over ET is signaled by a negative 
$G_{21}$. 

\subsection{Local transport}
\begin{figure}[h!]
 \includegraphics[width=8.75cm]{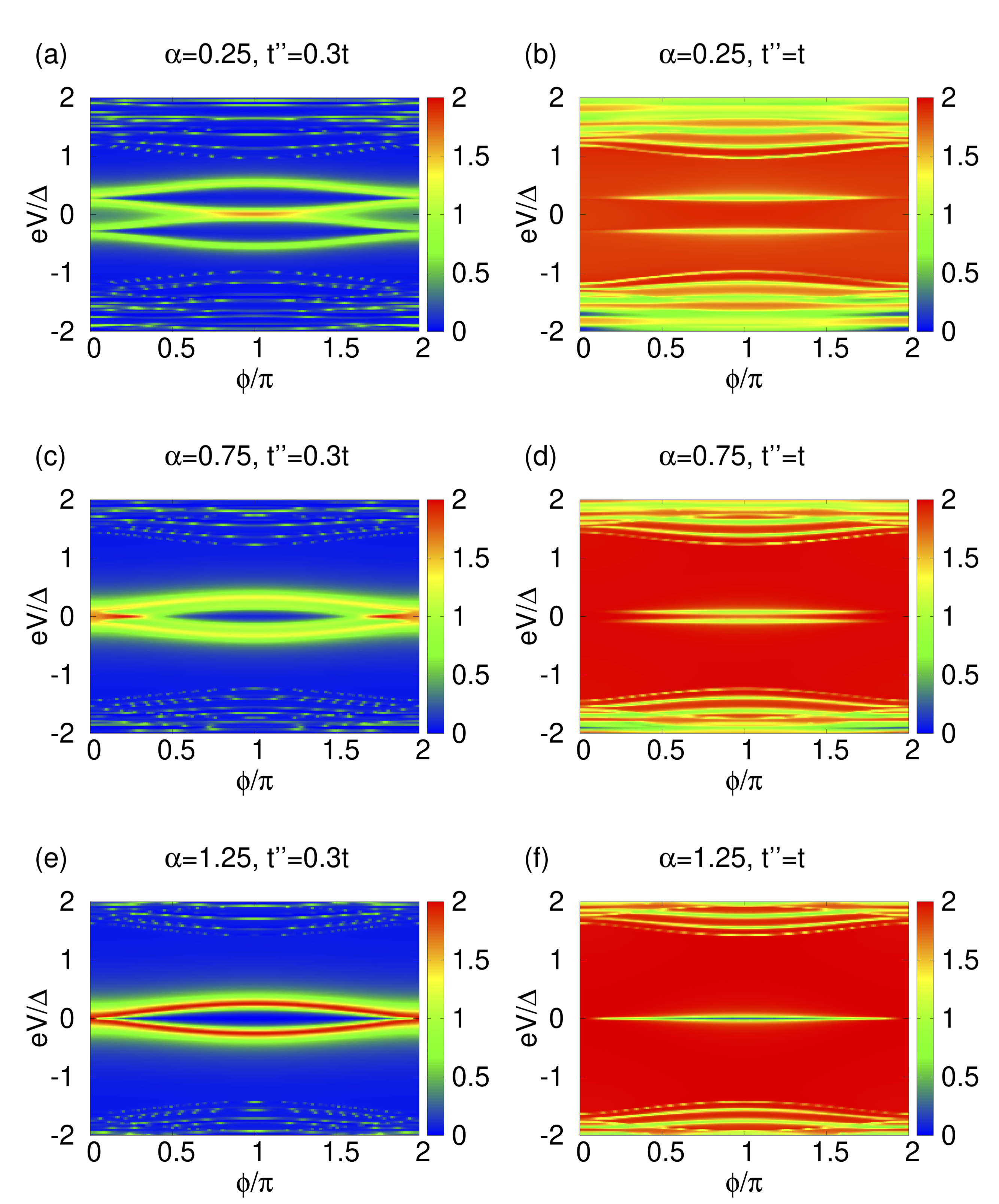}
\caption{The local conductance $G_{11}$ in units of $e^2/h$ as a
  function of bias voltage $V$ and the superconducting phase
  difference ($\phi$) with connection strength (a,c,e) $t''=0.3t$,
  (b,d,f) $t''=t$, for system parameters $\De=0.4t$, $t'=0.1t$,
  $\mu=0$ and $L=40$. The long range parameter is taken to be (a,b) $\alpha=0.25$, (c,d) $\alpha=0.75$ and (e,f) $\alpha=1.25$. (a,c,e)
  The peak in the local conductance captures the Majorana modes formed
  by the hybridization of MFs on two legs of the ladder.  (b,d,f) The
  local conductance is unable to provide sharp signatures for the
  Majorana modes due to the strong coupling between metal leads and
  superconductor. (a,b) Further, the long-range pairing ($\alpha$ small)
  hybridizes the Majorana edge modes which destroy their edge localization
  in the long-range limit $\alpha<<1$. (e,f) The system retains its
  Majorana edge modes on increasing $\alpha>>1$ which corresponds to the
  short-range limit of the Kitaev ladder.}
\label{fig:04}
\end{figure}

Now we turn to the conductance results when the ladder is connected to
the two normal metals and a bias is applied from $N_1$ grounding the
ladder and $N_2$. In Fig.~\ref{fig:04}, the variation of the
local conductance ($G_{11}$) as a function of the bias voltage $eV$
and the phase difference ($\phi$) is shown when both the normal metal
leads are connected to the Kitaev ladder with coupling strength
$t''$. First up, we observe that signatures of the Majorana modes are
masked in Fig.~\ref{fig:04}(b,d,f) whereas they are
clearly visible in Fig.~\ref{fig:04}(a,c,e). This indicates that
as the coupling $t''$ between the leads and the system is increased, 
perfect AR happens at most energies and the energy of the Andreev bound states  is not special.
In the short-range limit strongly dominating AR ($G_{11}\sim 2e^2/h$)
along the lines $eV=\pm\tilde{t}\sin{\frac{\phi}{2}}$ within the bound
$|eV|<\De$ as depicted in Fig.~\ref{fig:04}(e) characterize the
effect of the ABSs formed at the ends of the ladder. One can
see a direct resemblance between Fig.~\ref{fig:02}(e,f,g) and
Fig.~\ref{fig:04}(a,c,e) respectively. For all the
parameters in Fig.~\ref{fig:04} we see that $G_{11}$ tends to be
relatively small in the region $|eV|>\De$ due to presence of the
subgap Andreev states. Long-range pairings (as $\alpha$ is decreased) 
tend to destroy the localization of the MFs and ABSs in the system. The
extension of the Andreev bound states into the bulk of the system, in turn suppresses the local transport in the system as depicted in Fig.~\ref{fig:04}(a,c). 
Moreover, as $\alpha$ is decreased the lifting of the degeneracy between
the Majorana modes is evident, as was also clear from the energy spectrum 
shown in Fig.~\ref{fig:02}~(a,b). Though the local conductance $G_{11}$ in
Fig.~\ref{fig:04}(a,c) is peaked in the energy range of the Andreev bound states, the
value is below $e^2/h$ making it doubtful whether AR happens or not.
\begin{figure}[h!]
 \includegraphics[width=8.75cm]{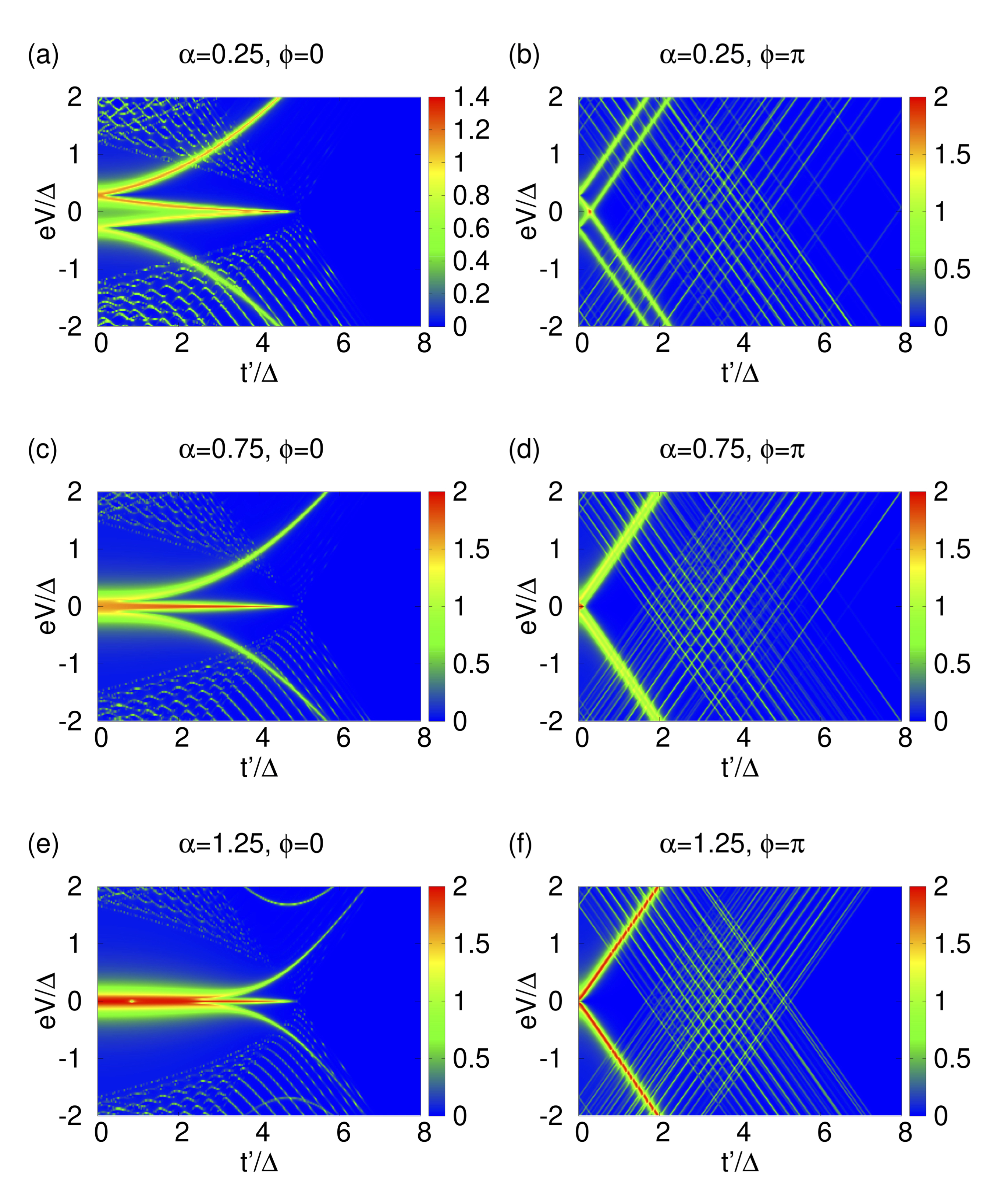}
\caption{The local conductance $G_{11}$ in units of $e^2/h$ as a function of  
bias voltage $V$ and the inter-leg hopping $t'$ with superconducting phase difference (a,c,e)$\phi=0$, (b,d,h) $\phi=\pi$ for the choice of system parameters
$\De=0.4t$, $t''=0.3t$, $\mu=0$ and $L=40$ along with the long-range parameters (a,b) $\alpha=0.25$, (c,d) $\alpha=0.75$ and (e,f) $\alpha=1.25$.}
\label{fig:05}
\end{figure}

Fig.~\ref{fig:02}(h) shows that when a non-zero $\phi$ is present for
a finite hopping $t'$, a lifting of the four-fold degenerate Majorana states by the formation of Andreev bound states is observed.
Long-range pairing further lifts the remaining degeneracy, as we have 
already observed in Fig.~\ref{fig:02}(e) and Fig.~\ref{fig:04}(a).
It is instructive to study the dependence of the local conductance as a
function of the inter-wire hopping $t'$.  In Fig.~\ref{fig:05}
the behavior of local conductance as a function of bias voltage $V$
and the inter-wire hopping $t'$ with two different values of $\phi$ and
for a range of values $\alpha$ is depicted. The local conductance for
$\phi=0$ in Fig.~\ref{fig:05}(a,c,e) shows close resemblance to
the energy spectrum of the isolated Kitaev ladder shown in
Fig.~\ref{fig:02}(a-c). We find that along the green lines for
$\alpha=0.25$, the local conductance in Fig.~\ref{fig:05}(a)
is very close to $0.8e^2/h$ implying a possible lack of AR. But on making the
system less long-ranged with $\alpha=0.75$ the local conductance($G_{11}$)
is enhanced to a value $\sim1.25e^2/h$ as shown in Fig.~\ref{fig:05}(c)
confirming a dominant AR. As the system tends to become more and more
short-ranged, the Majorana modes remain degenerate for small values of $t'$ 
as shown in Fig.~\ref{fig:05}(e). Thus,
the local conductance in the limit of small $t'$ and large $\alpha$ shows
strong AR with a conductance
value $2e^2/h$. For higher inter-leg coupling strength the system is still
not able to maintain perfectly localized Majorana modes and local
conductance also features lack of perfect AR as shown in
Fig.~\ref{fig:05}(e). For $\phi=\pi$, at finite $t'$ the MFs in the two 
legs of the ladder
hybridize to form Andreev bound states  and a finite $\alpha$ couples the ABSs at two ends of the 
ladder. In the short-range limit these Andreev bound states cause a split in the degenerate energy by an amount $\sim 2 t'/\Delta$ which is clearly captured by local transport
in Fig.~\ref{fig:05}(f). For $\phi=\pi$, the ladder dispersion
becomes $E=\pm(\sqrt{\ep_k^2+|\Delta_k|^2}+\nu t')$, where $\nu=\pm
1$. In the case of short-range pairing, $\Delta_k=2\Delta\sin{ka}$. 
So, for $t'\gtrsim2\De$, the gap closes and hence the features in
the region $t'\gtrsim2\De$ are due to subgap Andreev states as shown in
Fig.~\ref{fig:05}(b,d,f). Further, the lifting of degeneracy of the ABSs when
the pairing becomes long-ranged is captured nicely by local conductance ($G_{11}$)
as depicted in Fig.~\ref{fig:05}(b,d).

\begin{figure}[h!]
 \includegraphics[width=8.75cm]{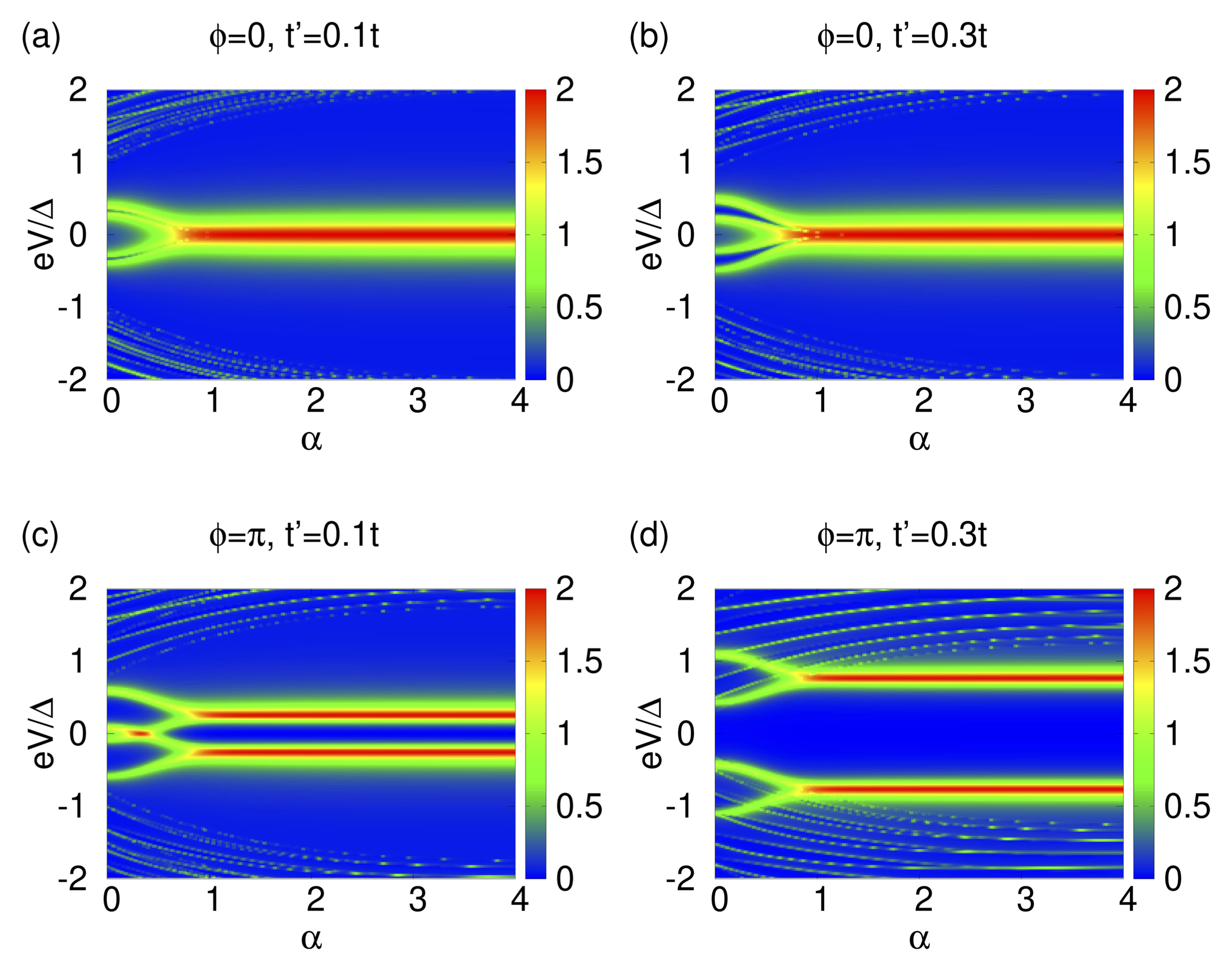}
 \caption{The local conductance $G_{11}$ in units of $e^2/h$ as a function of  
bias voltage $V$ and the long-range parameter $\alpha$ with  (a,b) $\phi=0$, (c,d) $\phi=\pi$ for other system parameters
$\De=0.4t$, $t''=0.3t$, $\mu=0$ and $L=40$ along with inter-leg hopping (a,c) $t'=0.1t$ and (b,d) $t'=0.3t$. The effective short range limit shifts further away from $\alpha=1$ for strong inter-leg coupling strength. }
\label{fig:06}
\end{figure}
The long-range pairing results in hybridized non-degenerate MFs and ABSs
at the two ends of the ladder. Here we look closely at
variation of local conductance as a function of the long-range parameter
$\alpha$. The transport characteristics of the Kitaev ladder with
various $\alpha$ is depicted in Fig.~\ref{fig:06}. For small
inter-leg hopping, short-range behavior seems to be obtained for
$\alpha \gtrsim 1$ as shown in Fig.~\ref{fig:06}(a). This is
in line with the results obtained for the long-range Kitaev 
chain~\citep{degottardi2011topological}.  As the inter-leg hopping 
$t'$ is increased, a greater critical
value of $\alpha$ seems to be needed before the system displays
short-range behavior as shown in Fig.~\ref{fig:06}(b). In
Fig.~\ref{fig:06}(a) the system shows a break in degeneracy
of Majorana modes for $\alpha<1$ as the local conductance attains a
value $\sim 0.7e^2/h$. But for $\alpha>1$ the extreme value $2e^2/h$ of local
conductance shows the strong presence of AR. Again, in the presence of strong
inter-leg hopping $t'$ we see that the critical value $\alpha_c$ above which degeneracy
remains intact is shifted to a greater value as shown in
Fig.~\ref{fig:06}(b).
\begin{table}[h!]
\begin{center}
\begin{tabular}{|c|c|c|c|}
\hline
$\textbf{S.}\atop\textbf{No.}$&\large$\mathbf{\alpha}$ & $\textbf{Degeneracy~} \atop \textbf{of~MFs}$ & $\textbf{Local}\atop \textbf{Conductance}$\\\hline
1.&$\alpha\lesssim 0.5$ & 0 & $\sim 0.7e^2/h$(low AR)\\
2.&$0.5\lesssim\alpha\lesssim1.25$ & 2 & $\sim1.4e^2/h$(medium AR)\\
3.&$\alpha\gtrsim1.25$ & 4 & $\sim 2e^2/h$(high AR)\\\hline
\end{tabular}
\end{center}
\caption{Different degenerate states of the system for different long-range parameter $\alpha$ with $\phi=0, t'=0.3t$.}
\label{table}
\end{table}
For $t'= 0.3 t$ the system possesses three
phases discussed in Table~\ref{table}. 
For a nonzero phase difference $\phi$ in the extreme short-range limit $\alpha\rightarrow\infty$ the presence of the inter-leg hopping is itself
sufficient to partially lift the four-fold degeneracy of the Majorana
modes, making them two-fold degenerate corresponding to each end of
the ladder.
\begin{figure}[h!]
\includegraphics[width=8.75cm]{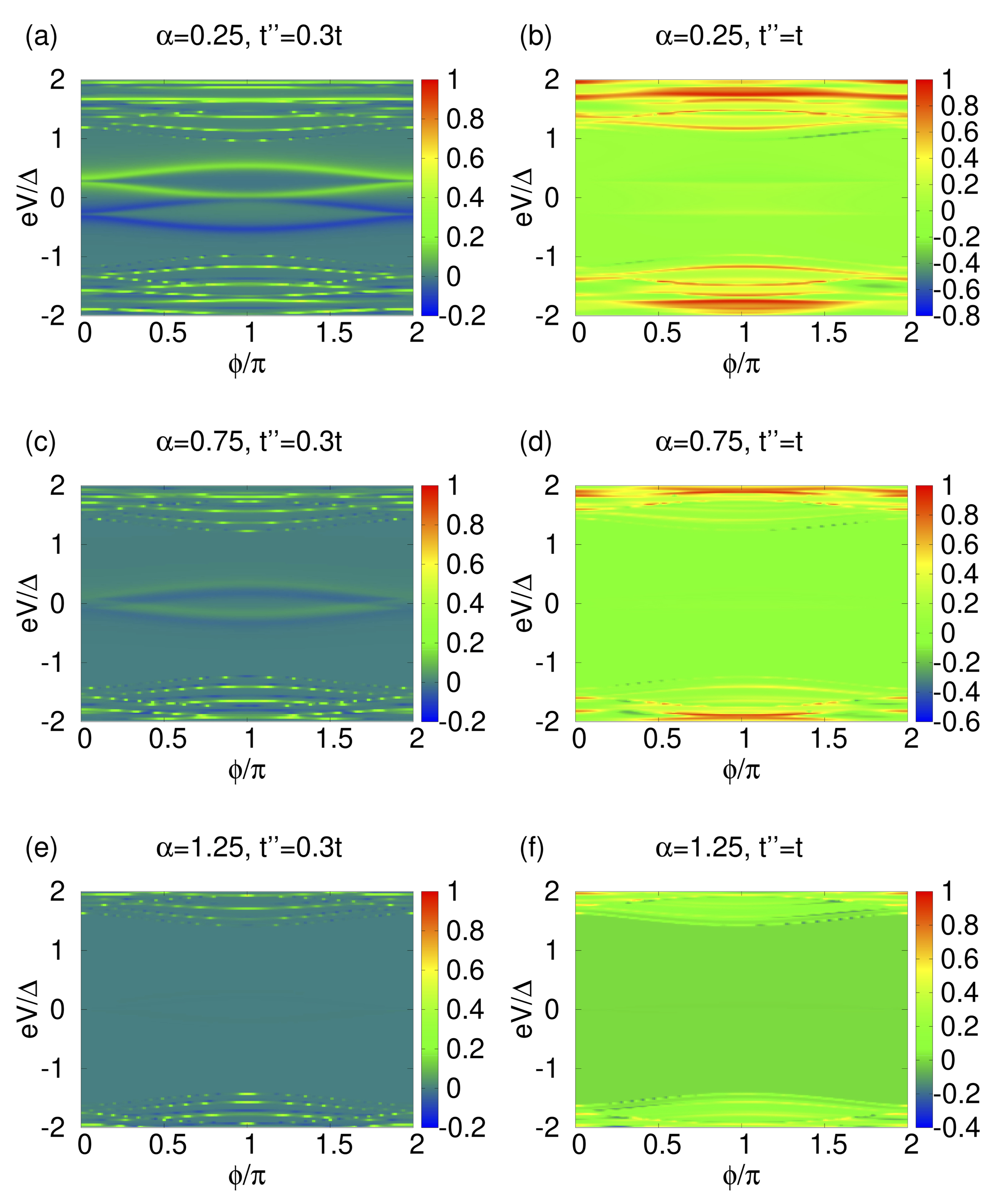}
 \caption{The transconductance $G_{21}$ in units of $e^2/h$ as a function of  
bias voltage $V$ and the superconducting phase difference $\phi$ for system parameters $\De=0.4t$, $\mu=0$, $t'=0.1t$, $L=40$ and (a,b) $\alpha=0.25$, (c,d) $\alpha=0.75$, (e,f) $\alpha=1.25$ along with (a,c,e) $t''=0.3t$ and (b,d,f) $t''=t$.}
\label{fig:07}
\end{figure}
 The ABSs formed are split in energy which can be seen in
Fig.~\ref{fig:02}(h), as is also captured by local conductance in
Fig.~\ref{fig:06}(c,d). Further we see from Fig.~\ref{fig:06}(c,d) that 
a large inter-leg hopping facilitates the full lifting of degeneracy of the ABSs in a much more effective way, in the long-range pairing regime.
\subsection{Nonlocal transport}
Before we conclude, we include a short
 discussion of non-local
transport with the help of a study of transconductance ($G_{21}$) in
this system.
 The presence of the highly localized Majorana fermions  and Andreev bound states  suppresses non-local transport. 
 On the other hand, the long-range pairing ($\alpha$ small) enhances the decay length of Majorana fermions(MFs) 
 and Andreev bound states in the Kitaev ladder, thus aiding non-local transport. Non-local transport is greatly
aided by a strong coupling between the normal metals and
superconductor~\cite{nehra19}, so for this part of the study we fix
$t''=t$.

\begin{figure}[h!]
\includegraphics[width=8.75cm]{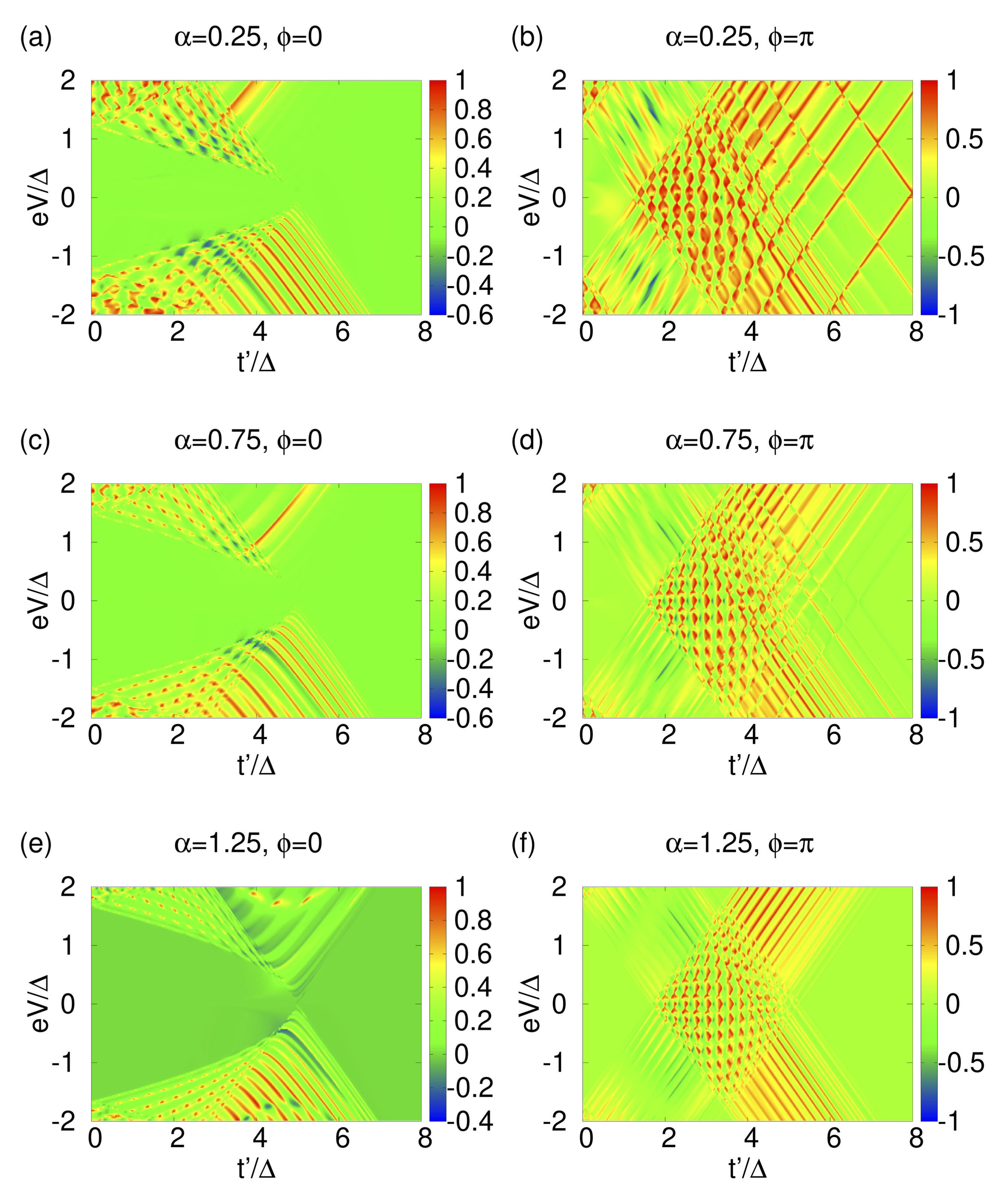}
\caption{The transconductance $G_{21}$ in units of $e^2/h$ as a function of  
bias voltage $V$ and the inter-leg hopping $t'$ for system parameters $\De=0.4t$, $\mu=0$, 
$t''=t$, $L=40$ and (a,c,e) $\phi=0$, (b,d,f) $\phi=\pi$. The long-range parameter varies as (a,b) $\alpha=0.25$, (c,d) $\alpha=0.75$ and (e,f) $\alpha=1.25$. The subgap Andreev states mediate the electron tunneling and crossed Andreev reflection through the Kitaev ladder.}
\label{fig:08}
\end{figure}
 The dispersion spectrum in Fig.~\ref{fig:02}(e-h) shows that
the long-range pairing causes the subgap Andreev states to be squeezed
within the superconducting gap region.
These subgap Andreev states are
responsible for non-local transport in the
system~\citep{nehra19,soori17}. 
Fig.~\ref{fig:07} studies transconductance
as a function of bias and the phase difference for different $\alpha$ and $t''$. 
From Fig.~\ref{fig:07}~(a) it can be seen that a pair of nonlocal Andreev bound states mediate CAR
which dominates over electron tunneling (ET). Also, the long-range pairing favors negative values of transconductance though
small in magnitude. In Fig.~\ref{fig:08}, the behavior of
transconductance is studied as a function of inter-leg hopping $t'$
and bias voltage $V$. The enhancement in CAR and ET are mediated by
subgap Andreev states near the boundary of the biasing window.  It is
interesting to note from Fig.~\ref{fig:08}(a,c,e) that long-range 
pairing can enhance CAR over ET making the transconductance
negative even in the absence of superconducting phase difference
$\phi$~\cite{nehra19}. The presence of superconducting phase $\phi$
further shows dominance of CAR over ET in
Fig.~\ref{fig:08}(b,d,f) along the energies of Andreev bound
states.
\begin{figure}
\includegraphics[width=8.75cm]{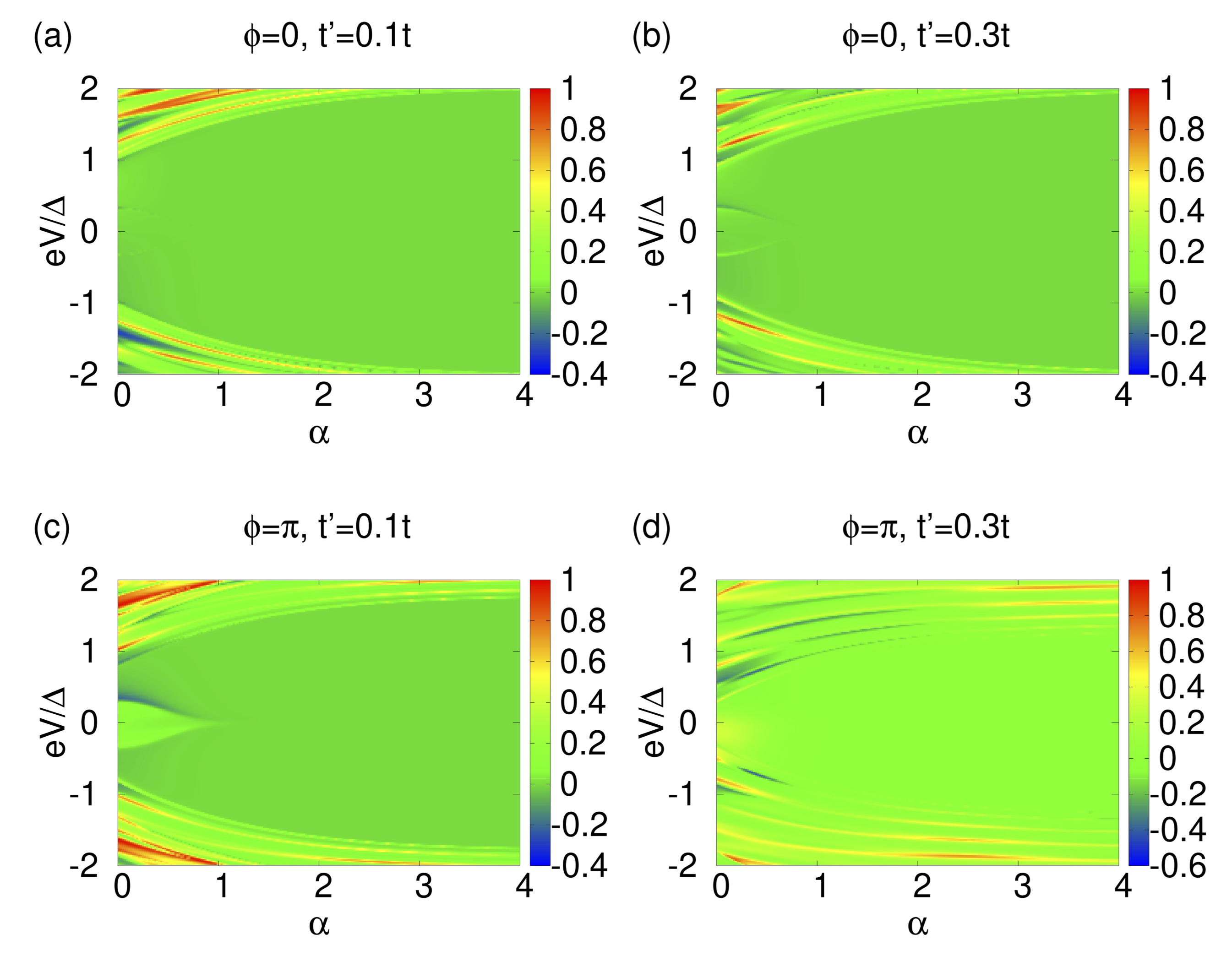}
 \caption{The transconductance $G_{21}$ in units of $e^2/h$ as a function of  
bias voltage $V$ and the long-range parameter $\alpha$ for system parameters
$\De=0.4t$, $t''=t$, $\mu=0$, $L=40$ and (a,c) $t'=0.1t$, (b,d) $t'=0.3t$ for (a,b) $\phi=0$, (c,d) $\phi=\pi$.}
\label{fig:09}
\end{figure}
 Fig.~\ref{fig:09} features the transconductance
($G_{21}$) for a range of $\alpha$ and bias voltage
$V$. It can be seen that an enhancement of CAR with nearly extreme
values $\sim -0.6 e^2/h$ of transconductance at some special points is
possible in the long-range limits along with superconducting phase
$\phi$ and inter-leg hopping $t'$.

\section{Summary}
In this paper, we have studied the connection between the equilibrium
and topological properties of a long-range Kitaev ladder and transport
through it when the system is connected to metallic leads.  Depending
on the value of the long-range parameter $\alpha$, the system becomes
either short-ranged or long-ranged. The superconducting phase difference breaks different
symmetries in the system resulting in a change of the topological
class of the Kitaev long-range ladder; however, topological properties
are unaffected by the long-range nature of the pairing. Next, we
studied local and nonlocal transport through the Kitaev ladder, when it
is connected to two normal metals. The isolated long-range Kitaev
ladder hosts Andreev bound states formed by a
recombination of Majorana end modes and standing waves formed by a 
recombination of subgap Andreev states. These states mediate local and
nonlocal transport respectively. The local conductance is enhanced except for
certain parameter choices signaling a dominant Andreev reflection in
the short-range limit. Andreev reflection is suppressed by long-range pairing
(when $\alpha$ is small) and inter-wire hopping. We also find that long-range pairing
alone can enhance crossed Andreev reflection without the need of a superconducting 
phase difference proposed earlier~\cite{nehra19,soori17}. For weak normal metal Kitaev
ladder hopping, the dependence of local conductance on superconducting phase
difference shows the splitting of Majorana state energies and the
bands of subgap Andreev states.  Thus, we have studied the effect of
Andreev bound states formed by the hybridization of Majorana fermions
and subgap Andreev states on local and nonlocal transport in the
normal metal-Kitaev ladder-normal metal system. While the enhancement
of Andreev reflection due to Majorana states is well known, we show
that these states can enhance nonlocal transport when the metal-ladder interface hopping is strong though the magnitude of
transconductance is low. For strong inter-wire hopping, nonlocal conductance (mediated by subgap Andreev
states) is strongly enhanced for certain
choices of the parameters.  

\acknowledgements This work has benefited from
discussion with Surajit Sarkar. Auditya Sharma is grateful for financial support via the DST-INSPIRE
Faculty Award [DST/INSPIRE/04/2014/002461]. Abhiram Soori thanks
DST-INSPIRE Faculty Award (Faculty Reg. No.~:~IFA17-PH190) for
financial support.  

\bibliography{ref_mfladder} 

\appendix
\section{Equation of motion at the boundaries}
The appendix contains the detailed calculation of the various scattering 
amplitudes of all four processes. The wavefunction in two normal metal 
regions on $j^{th}$ site is given by $[\psi^e_j,~\psi^h_j]^T$ where
\begin{align}
\psi^e_{j}=&e^{ik_eaj}+r_e e^{-ik_eaj}\quad\text{for}\quad j\le 0\\
\psi^h_{j}=&r_he^{ik_haj}\quad \text{for}\quad j\le 0\\
\psi^e_{j}=&t_e e^{ik_eaj}\quad\text{for}\quad j\ge L+1\\
\psi^h_{j}=&t_h e^{-ik_haj}\quad\text{for}\quad j\ge L+1
\end{align}
Here, $r_e$, $r_h$ are the reflection coefficients and $t_e$, $t_h$ are
the transmission coefficients of electrons and holes in the two metallic 
leads. Also, momenta of electron and hole are given by 
$k_ea=\cos^{-1}\big(\frac{E+\mu}{2t}\big)$ and 
$k_ha=\cos^{-1}\big(\frac{E-\mu}{2t}\big)$ respectively. In the ladder 
region, the wave function has the form 
$[\psi^e_{j,\sigma},\psi^h_{j,\sigma}]^T$ with $1\leq j\leq L$ as 
the site index and $\sigma=1,2$ as the ladder leg index. Here, $a$ is 
the lattice spacing.
There are total $4L+4$ unknowns i.e. wavefunctions in ladder region and
4 scattering coefficients. Also, for a given energy we have $4L+4$ 
equation of motion at boundaries and in ladder region. These
unknowns can be calculated by writing the equations of motion from the
Hamiltonian (Eq.~\ref{eq:fullham}) at the two boundaries and in ladder
region. These equations of motion at the left metal - ladder junction
are given by:
\begin{align}
&E\psi_{0}^{e}=-t''\psi_{1}^{e}-t\psi_{-1}^{e}-\mu\psi_{0}^{e}\\
&E\psi_{0}^{h}=t''\psi_{1}^{h}+t\psi_{-1}^{h}+\mu\psi_{0}^{h}.
\end{align}
The equations of motion for upper leg($\sigma=1$) of the long-range Kitaev 
ladder at a given energy $E$ are
\begin{widetext}
\begin{align}
&E\psi_{1,1}^{e}=-t\psi_{2,1}^{e}-t''\psi_{0}^{e}-\mu\psi_{1,1}^{e}-t^{\prime}\psi_{1,2}^{e}-\displaystyle\sum_{j'=2}^{L}\frac{\Delta e^{i\phi_1}}{(j'-1)^{\alpha}}\psi_{j',1}^{h}\\
&E\psi_{2,1}^{e}=-t(\psi_{1,1}^{e}+\psi_{3,1}^{e})-\mu\psi_{2,1}^{e}-t^{\prime}\psi_{2,2}^{e}-\displaystyle\sum_{j'=3}^{L}\frac{\Delta e^{i\phi_1}}{(j'-2)^{\alpha}}\psi_{j',1}^{h}+\Delta e^{i\phi_1} \psi_{1,1}^{h}\\
&\hspace{20pt}\vdots\hspace{80pt}\vdots\hspace{80pt}\vdots\nonumber\\
&E\psi_{L-1,1}^{e}=-t(\psi_{L-2,1}^{e}+\psi_{L,1}^{e})-\mu\psi_{L-1,1}^{e}-t^{\prime}\psi_{L-1,2}^{e}+\displaystyle\sum_{j'=1}^{L-2}\frac{\Delta e^{i\phi_1}}{(L-1-j')^{\alpha}}\psi_{j',1}^{h}- \Delta e^{i\phi_1} \psi_{L,1}^{h}\\
&E\psi_{L,1}^{e}=-t''\psi_{L+1}^{e}-t\psi_{L-1,1}^{e}-\mu\psi_{L,1}^{e}-t^{\prime}\psi_{L,2}^{e}+\displaystyle\sum_{j'=1}^{L-1}\frac{\Delta e^{i\phi_1}}{(L-j')^{\alpha}}\psi_{j',1}^{h}\\
&E\psi_{1,1}^{h}=t\psi_{2,1}^{h}+t''\psi_{0}^{h}+\mu\psi_{1,1}^{h}+t^{\prime}\psi_{1,2}^{h}+\displaystyle\sum_{j'=2}^{L}\frac{\Delta e^{-i\phi_1}}{(j'-1)^{\alpha}}\psi_{j',1}^{e}\\
&E\psi_{2,1}^{h}=t(\psi_{1,1}^{h}+\psi_{3,1}^{h})+\mu\psi_{2,1}^{h}+t^{\prime}\psi_{2,2}^{h}+\displaystyle\sum_{j'=3}^{L}\frac{\Delta e^{-i\phi_1}}{(j'-2)^{\alpha}}\psi_{j',1}^{e}- \Delta e^{-i\phi_1} \psi_{1,1}^{e}\\
&\hspace{20pt}\vdots\hspace{80pt}\vdots\hspace{80pt}\vdots\nonumber\\
&E\psi_{L-1,1}^{h}=t(\psi_{L-2,1}^{h}+\psi_{L,1}^{h})+\mu\psi_{L-1,1}^{h}+t^{\prime}\psi_{L-1,2}^{h}-\displaystyle\sum_{j'=1}^{L-2}\frac{\Delta e^{-i\phi_1}}{(L-1-j')^{\alpha}}\psi_{j',1}^{e}+\Delta e^{-i\phi_1}\psi_{L,1}^{e}\\
&E\psi_{L,1}^{h}=t''\psi_{L+1}^{h}+t\psi_{L-1,1}^{h}+\mu\psi_{L,1}^{h}+t^{\prime}\psi_{L,2}^{h}-\displaystyle\sum_{j'=1}^{L-1}\frac{\Delta e^{-i\phi_1}}{(L-j')^{\alpha}}\psi_{j',1}^{e}.
\end{align}
\end{widetext}
The equations of motion for lower leg($\sigma=2$) of the long-range Kitaev ladder with incoming energy $E$ of the particle are
\begin{widetext}
\begin{align}
&E\psi_{1,2}^{e}=-t\psi_{2,2}^{e}-\mu\psi_{1,2}^{e}-t^{\prime}\psi_{1,1}^{e}-\displaystyle\sum_{j'=2}^{L}\frac{\Delta e^{i\phi_2}}{(j'-1)^{\alpha}}\psi_{j',2}^{h}\\
&E\psi_{2,2}^{e}=-t(\psi_{1,2}^{e}+\psi_{3,2}^{e})-\mu\psi_{2,2}^{e}-t^{\prime}\psi_{2,1}^{e}-\displaystyle\sum_{j'=3}^{L}\frac{\Delta e^{i\phi_2}}{(j'-2)^{\alpha}}\psi_{j',2}^{h}+\Delta e^{i\phi_2}\psi_{1,2}^{h}\\
&\hspace{20pt}\vdots\hspace{80pt}\vdots\hspace{80pt}\vdots\\
&E\psi_{L-1,2}^{e}=-t(\psi_{L-2,2}^{e}+\psi_{L,2}^{e})-\mu\psi_{L-1,2}^{e}-t^{\prime}\psi_{L-1,1}^{e}+\displaystyle\sum_{j'=1}^{L-2}\frac{\Delta e^{i\phi_2}}{(L-1-j')^{\alpha}}\psi_{j',2}^{h}- \Delta e^{i\phi_2} \psi_{L,2}^{h}\\
&E\psi_{L,2}^{e}=-t\psi_{L-1,2}^{e}-\mu\psi_{L,2}^{e}-t^{\prime}\psi_{L,1}^{e}+\displaystyle\sum_{j'=1}^{L-1}\frac{\Delta e^{i\phi_2}}{(L-j')^{\alpha}}\psi_{j',2}^{h}\\
&E\psi_{1,2}^{h}=t\psi_{2,2}^{h}+\mu\psi_{1,2}^{h}+t^{\prime}\psi_{1,1}^{h}+\displaystyle\sum_{j'=2}^{L}\frac{\Delta e^{-i\phi_2}}{(j'-1)^{\alpha}}\psi_{j',2}^{e}\\
&E\psi_{2,2}^{h}=t(\psi_{1,2}^{h}+\psi_{3,2}^{h})+\mu\psi_{2,2}^{h}+t^{\prime}\psi_{2,1}^{h}+\displaystyle\sum_{j'=3}^{L}\frac{\Delta e^{-i\phi_2}}{(j'-1)^{\alpha}}\psi_{j',2}^{e}-\Delta e^{-i\phi_2}\psi_{1,2}^{e}\\
&\hspace{20pt}\vdots\hspace{80pt}\vdots\hspace{80pt}\vdots\nonumber\\
&E\psi_{L-1,2}^{h}=t(\psi_{L-2,2}^{h}+\psi_{L,2}^{h})+\mu\psi_{L-1,2}^{h}+t^{\prime}\psi_{L-1,1}^{h}-\displaystyle\sum_{j'=1}^{L-2}\frac{\Delta e^{-i\phi_2}}{(L-1-j')^{\alpha}}\psi_{j',2}^{e}+ \Delta e^{-i\phi_2} \psi_{L,2}^{e}\\
&E\psi_{L,2}^{h}=t\psi_{L-1,2}^{h}+\mu\psi_{L,2}^{h}+t^{\prime}\psi_{L,1}^{h}-\displaystyle\sum_{j'=1}^{L-1}\frac{\Delta e^{-i\phi_2}}{(L-j')^{\alpha}}\psi_{j',2}^{e}.
\end{align}
\end{widetext}
The equations of motion at the right normal metal - ladder junction for a given energy $E$ are:
\begin{align}
&E\psi_{L+1}^{e}=-t\psi_{L+2}^{e}-t''\psi_{L}^{e}-\mu\psi_{L+1}^{e}\\
&E\psi_{L+1}^{h}=t\psi_{L+2}^{h}+t''\psi_{L}^{h}+\mu\psi_{L+1}^{h}.
\end{align}
Different unknowns can be computed by solving these equations. Finally, the conductances can be then calculated from Eq.~\ref{eq:conductance}.  

\end{document}